\documentclass[11pt]{article}

\usepackage{amsfonts,amssymb,chet,dsfont,graphicx,mathrsfs,pgfplots,tikz}
\allowdisplaybreaks

\newcommand{\1}{\mathds{1}}

\newcommand{\Op}[2]{\mathcal{O}_{#1}(\eta_{#2})}

\newcommand{\ee}[3]{(\eta_{#1}\cdot\eta_{#2})^{#3}}

\newcommand{\D}{\mathcal{D}}
\newcommand{\A}{\mathcal{A}}
\newcommand{\cOPE}[4]{{}_{#1}c_{#2#3}^{\phantom{#2#3}#4}}
\newcommand{\DOPE}[4]{{}_{#1}\D_{#2#3}^{\phantom{#2#3}#4}}
\newcommand{\tOPE}[6]{{}_{#1}t_{#2#3}^{#5#6#4}}

\newcommand{\Vev}[1]{\left\langle{#1}\right\rangle}

\title{Conformal Two-Point Correlation Functions\\from the Operator Product Expansion}

\author{Jean-Fran\c{c}ois Fortin$^{\ast,}$\email{jean-francois.fortin@phy.ulaval.ca}, Valentina Prilepina$^{\ast,}$\email{valentina.prilepina.1@ulaval.ca} and Witold Skiba$^{\dagger,}$\email{witold.skiba@yale.edu}}

\affiliation{
$^\ast$D\'epartement de Physique, de G\'enie Physique et d'Optique\\Universit\'e Laval, Qu\'ebec, QC G1V 0A6, Canada\\
$^\dagger$Department of Physics, Yale University, New Haven, CT 06520, USA
}

\abstract{We compute the most general embedding space two-point function in arbitrary Lorentz representations in the context of the recently introduced formalism in \cite{Fortin:2019fvx,Fortin:2019dnq}.  This work provides a first explicit application of this approach and furnishes a number of checks of the formalism.  We project the general embedding space two-point function to position space and find a form consistent with conformal covariance.  Several concrete examples are worked out in detail.  We also derive constraints on the OPE coefficient matrices appearing in the two-point function, which allow us to impose unitarity conditions on the two-point function coefficients for operators in any Lorentz representations.}

\date{June 2019} 

\begin{document}

\maketitle



\section{Introduction}\label{SecIntro}

Conformal field theories (CFTs) are special quantum field theories (QFTs) endowed with a powerful invariance under a broad group of symmetries, the conformal group $SO(2,d)$.  CFTs represent fixed points in renormalization group flows in the space of QFTs, describe second order phase transitions in statistical physics systems, and shed light on the universal structure of the landscape of all QFTs.  Moreover, they prescribe a non-perturbative approach for the investigation of quantum gravity theories via the AdS/CFT correspondence.

A modern revival of interest in the subject was initiated by \cite{Dolan:2000ut,Dolan:2003hv,Dolan:2011dv,Rattazzi:2008pe}.  In recent years, tremendous progress has been made in the exploration of CFTs, largely owing to the power of the conformal bootstrap, a systematic program of applying consistency conditions and crossing symmetry to carve out the space of allowed theories, an idea introduced in \cite{Ferrara:1973yt,Polyakov:1974gs}.  A review of modern bootstrap and further references can be found in \cite{Poland:2018epd}. The ultimate dream of this program is to classify all CFTs as relevant deformations of a small subset of CFTs in the spirit of the Jacobi identity.

A natural habitat for the formulation of CFTs is the embedding space, where the conformal transformations act linearly \cite{Dirac:1936fq}.  The organic observables in CFTs are correlation functions of quasi-primary operators.  A complete implementation of the bootstrap calls for the determination of the four-point conformal blocks for general Lorentz representations.  Recently, a novel formalism for a unified treatment of arbitrary $M$-point correlations functions in the embedding space was introduced in \cite{Fortin:2019fvx,Fortin:2019dnq}.  This framework lays out a program that enables the efficient computation of all possible blocks and further empowers the determination of all $M$-point functions.  It relies on a reinterpretation of the embedding space operator product expansion (OPE) using a new uplift for general quasi-primary operators.

The OPE constitutes replacing the product of two local operators at distinct spacetime points $x_1$ and $x_2$ by an infinite sum of operators at some point inside the interval $[x_1,x_2]$.  While in general the OPE converges only in the asymptotic short-distance limit, in a CFT it is known to converge at finite separation, owing to the enhanced symmetry of the theory~\cite{Mack:1976pa}.  The OPE is therefore a well-defined fundamental quantity in a CFT, where its utility lies in formulating $M$-point correlation functions in terms of $(M-1)$-point functions.  The OPE in a CFT expresses the product of two quasi-primary operators at non-coincident points in terms of a series in quasi-primary operators and their descendants.  Explicitly, the embedding space OPE can be stated as
\eqn{\Op{i}{1}\Op{j}{2}=\sum_k\sum_{a=1}^{N_{ijk}}\cOPE{a}{i}{j}{k}\DOPE{a}{i}{j}{k}(\eta_1,\eta_2)\Op{k}{2},}[EqOPE]
where the sum over quasi-primary operators $\Op{k}{2}$ is infinite, while the sum over the $N_{ijk}$ OPE coefficients $\cOPE{a}{i}{j}{k}$ with the appropriate differential operators $\DOPE{a}{i}{j}{k}(\eta_1,\eta_2)$ is finite.  The sum includes all linearly independent quasi-primary operators, while the infinite towers of descendants are accounted for by the differential operators $\DOPE{a}{i}{j}{k}(\eta_1,\eta_2)$.  In a CFT, the form of the two-point correlation functions is completely determined by symmetry considerations.  From the OPE \eqref{EqOPE} in Lorentzian signature, the two-point functions are
\eqn{\Vev{\Op{i}{1}\Op{j}{2}}=(\mathcal{T}_{12}^{\boldsymbol{N}}\Gamma)(\mathcal{T}_{21}^{\boldsymbol{N}^C}\Gamma)\cdot\frac{\lambda_{\boldsymbol{N}}\cOPE{}{i}{j}{\1}\hat{\mathcal{P}}_{12}^{\boldsymbol{N}}}{\ee{1}{2}{\tau}}.}[Eq2ptOPE]
Hence, once the OPE is known, so are technically all possible correlation functions, up to the OPE coefficients.

In this work, we give an explicit application of the program set forth in \cite{Fortin:2019fvx,Fortin:2019dnq}, using it to compute the general two-point function for generic Lorentz representations (see \cite{Rattazzi:2010gj,Poland:2010wg,Costa:2011mg,SimmonsDuffin:2012uy,Hartman:2016dxc,Hofman:2016awc,Hartman:2016lgu,Afkhami-Jeddi:2016ntf,Cuomo:2017wme} for various results on two- and three-point functions).  This is a useful first step in the successful implementation of the new framework.  In particular, the two-point correlator carries projection operators that encapsulate all the essential group theoretic information which subsequently feeds in to the computations of the three-point, four-point, and general $M$-point functions.

This paper is organized as follows: Section \ref{SecProj} discusses the projection operators which are ubiquitous in the construction of correlation functions.  Section \ref{Sec2ptES} describes two-point correlation functions in embedding space using the formalism developed in \cite{Fortin:2019fvx,Fortin:2019dnq}.  The required tensor structures are obtained in terms of the projection operators and their normalization is chosen.  In Section \ref{Sec2ptPS}, the two-point correlation functions are projected to position space for quasi-primary operators in general irreducible representations of the Lorentz group.  The resulting position space two-point correlation functions turn out to be the expected correlation functions obtained from the usual symmetry arguments, and these are shown to be covariant under conformal transformations by direct computation.  Some specific examples (with both fundamental and mixed representations) are also discussed.  This section provides a first explicit sanity check on the formalism introduced in \cite{Fortin:2019fvx,Fortin:2019dnq}.  For completeness, Section \ref{SecUnitarity} determines the unitarity conditions on the coefficients of two-point functions.  Finally, conclusions are presented in Section \ref{SecConc}.  Throughout this paper, we use the notation and conventions detailed in \cite{Fortin:2019dnq}.


\section{Hatted Projection Operators and Half-Projectors}\label{SecProj}

From the formalism developed in \cite{Fortin:2019fvx,Fortin:2019dnq}, non-vanishing two-point correlation functions for quasi-primary operators $\Op{i}{}$ and $\Op{j}{}$ are simply and suggestively given by \eqref{Eq2ptOPE} where $\boldsymbol{N}=\{N_1,\ldots,N_r\}$ denotes an irreducible Lorentz representation of the operator $\Op{i}{}$ and its corresponding Dynkin indices, while $\boldsymbol{N}^C$ denotes the conjugate representation.\footnote{See Section \ref{Sec2ptES} for a discussion of the relation between contragredient-reflected representations and conjugate representations in Lorentzian signature.}  We will discuss the details in the next section.

Of interest here is $\hat{\mathcal{P}}_{12}^{\boldsymbol{N}}$ which is an embedding space projection operator to the representation $\boldsymbol{N}$, while $(\mathcal{T}_{12}^{\boldsymbol{N}}\Gamma)$ is what we refer to as a half-projector.  It is evident from \eqref{Eq2ptOPE} that the particulars of the projection operators are central in the determination of two-point correlation functions of quasi-primary operators in general irreducible representations of the Lorentz group.  Hence, the projection operators are the focus of this section.

It is more instructive to discuss the projectors and half-projectors in position space first.  The corresponding quantities in the embedding space are directly related to the ones in position space.  The half-projectors serve to translate the spinor indices carried by each operator to the ``dummy" vector and spinor indices that need to be summed over when constructing correlation functions.  They earned their name because they square to form projection operators.  The hatted projectors operate on the dummy indices alone.

The position space projection operators satisfy the following essential properties: (1) the projection property $\hat{\mathcal{P}}^{\boldsymbol{N}}\cdot\hat{\mathcal{P}}^{\boldsymbol{N}'}=\delta_{\boldsymbol{N}'\boldsymbol{N}}\hat{\mathcal{P}}^{\boldsymbol{N}}$, (2) the completeness relation $\sum_{\boldsymbol{N}|n_v\text{ fixed}}\hat{\mathcal{P}}^{\boldsymbol{N}}=\1-\text{traces}$, and (3) the tracelessness condition $g\cdot\hat{\mathcal{P}}^{\boldsymbol{N}}=\gamma\cdot\hat{\mathcal{P}}^{\boldsymbol{N}}=\hat{\mathcal{P}}^{\boldsymbol{N}}\cdot g=\hat{\mathcal{P}}^{\boldsymbol{N}}\cdot\gamma=0$.  We next discuss some simple algorithms for the construction of hatted projection operators to general irreducible representations of the Lorentz group.


\subsection{Projection Operators for Defining Irreducible Representations}

Hatted projection operators to general irreducible representations can be built from the corresponding operators for the defining irreducible representations.  It is therefore appropriate to first describe the hatted projectors to defining irreducible representations.

In odd spacetime dimensions, the hatted projectors to the defining irreducible representations are
\eqn{
\begin{gathered}
(\hat{\mathcal{P}}^{\boldsymbol{e}_r})_\alpha^{\phantom{\alpha}\beta}=\delta_\alpha^{\phantom{\alpha}\beta},\qquad(\hat{\mathcal{P}}^{\boldsymbol{e}_{i\neq r}})_{\mu_i\cdots\mu_1}^{\phantom{\mu_i\cdots\mu_1}\nu_1\cdots\nu_i}=\delta_{[\mu_1}^{\phantom{[\mu_1}\nu_1}\cdots\delta_{\mu_i]}^{\phantom{\mu_i]}\nu_i},\qquad(\hat{\mathcal{P}}^{2\boldsymbol{e}_r})_{\mu_r\cdots\mu_1}^{\phantom{\mu_r\cdots\mu_1}\nu_1\cdots\nu_r}=\delta_{[\mu_1}^{\phantom{[\mu_1}\nu_1}\cdots\delta_{\mu_r]}^{\phantom{\mu_r]}\nu_r},
\end{gathered}
}[EqPodd]
while in even dimensions they are given by
\eqn{
\begin{gathered}
(\hat{\mathcal{P}}^{\boldsymbol{e}_{r-1}})_\alpha^{\phantom{\alpha}\beta}=\delta_\alpha^{\phantom{\alpha}\beta},\qquad(\hat{\mathcal{P}}^{\boldsymbol{e}_r})_{\tilde{\alpha}}^{\phantom{\tilde{\alpha}}\tilde{\beta}}=\delta_{\tilde{\alpha}}^{\phantom{\tilde{\alpha}}\tilde{\beta}},\qquad(\hat{\mathcal{P}}^{\boldsymbol{e}_{i\neq r-1,r}})_{\mu_i\cdots\mu_1}^{\phantom{\mu_i\cdots\mu_1}\nu_1\cdots\nu_i}=\delta_{[\mu_1}^{\phantom{[\mu_1}\nu_1}\cdots\delta_{\mu_i]}^{\phantom{\mu_i]}\nu_i},\\
(\hat{\mathcal{P}}^{\boldsymbol{e}_{r-1}+\boldsymbol{e}_r})_{\mu_{r-1}\cdots\mu_1}^{\phantom{\mu_{r-1}\cdots\mu_1}\nu_1\cdots\nu_{r-1}}=\delta_{[\mu_1}^{\phantom{[\mu_1}\nu_1}\cdots\delta_{\mu_{r-1}]}^{\phantom{\mu_{r-1}]}\nu_{r-1}},\\
(\hat{\mathcal{P}}^{2\boldsymbol{e}_{r-1}})_{\mu_r\cdots\mu_1}^{\phantom{\mu_r\cdots\mu_1}\nu_1\cdots\nu_r}=\frac{1}{2}\delta_{[\mu_1}^{\phantom{[\mu_1}\nu_1}\cdots\delta_{\mu_r]}^{\phantom{\mu_r]}\nu_r}+(-1)^r\frac{\mathscr{K}}{2r!}\epsilon_{\mu_1\cdots\mu_r}^{\phantom{\mu_1\cdots\mu_r}\nu_r\cdots\nu_1},\\
(\hat{\mathcal{P}}^{2\boldsymbol{e}_r})_{\mu_r\cdots\mu_1}^{\phantom{\mu_r\cdots\mu_1}\nu_1\cdots\nu_r}=\frac{1}{2}\delta_{[\mu_1}^{\phantom{[\mu_1}\nu_1}\cdots\delta_{\mu_r]}^{\phantom{\mu_r]}\nu_r}-(-1)^r\frac{\mathscr{K}}{2r!}\epsilon_{\mu_1\cdots\mu_r}^{\phantom{\mu_1\cdots\mu_r}\nu_r\cdots\nu_1}.
\end{gathered}
}[EqPeven]
In \eqref{EqPodd} and \eqref{EqPeven}, $\delta_{[\mu_1}^{\phantom{[\mu_1}\nu_1}\cdots\delta_{\mu_i]}^{\phantom{\mu_i]}\nu_i}$ is the totally antisymmetric normalized product of $\delta_\mu^{\phantom{\mu}\nu}$, while $\mathscr{K}$ is the proportionality constant in $\gamma^{\mu_1\cdots\mu_d}=\mathscr{K}\epsilon^{\mu_1\cdots\mu_d}\1$ which satisfies $\mathscr{K}^2=(-1)^{r+1}$ with $\epsilon^{1\cdots d}=1$.  From \eqref{EqPodd} and \eqref{EqPeven}, it is straightforward to verify that these operators indeed satisfy the requisite projection property and  tracelessness condition.  The completeness relation can subsequently be used to generate other hatted projectors.


\subsection{Projection Operators for Arbitrary Irreducible Representations}

An arbitrary irreducible representation of $SO(p,q)$ is indexed by a set of non-negative integers, the Dynkin indices, denoted by $\boldsymbol{N}=\{N_1,\ldots,N_r\}\equiv\sum_iN_i\boldsymbol{e}_i$, where $r$ is the rank of the Lorentz group and $\boldsymbol{e}_i$ is the usual unit vector $\boldsymbol{e}_i\equiv(\boldsymbol{e}_i)_j=\delta_{ij}$.  Clearly, the defining representations are special cases of general irreducible representations.

There exist several techniques for the construction of hatted projection operators to general irreducible representations of the Lorentz group.  For example, we can resort to Young tableaux techniques with the birdtrack notation \cite{Cvitanovic:2008zz,Costa:2014rya,Costa:2016hju} as well as the weight-shifting formalism \cite{Karateev:2017jgd,Costa:2018mcg}.  Whatever the approach, the construction of the hatted projection operators amounts to an exercise in group theory, and the path used to obtain them is irrelevant; only the final result is of importance here.

Another construction technique is based on the tensor product decomposition, the projection property, the completeness relation, and the tracelessness condition (see \textit{e.g.} \cite{Fortin:2019dnq}).  Although not particularly efficient, it can in principle be used to generate the projector to any irreducible representation via recursion, including the general spinorial representations, which constitute a limitation for other methods.  The technique can be appropriately adapted so as to render it more efficient following \cite{Rejon-Barrera:2015bpa}.  For simplicity, we restrict the discussion here to odd spacetime dimensions, as there is only one defining spinor representation to consider in that case.  The generalization to even dimensions is straightforward.

We can construct the hatted projection operator to a general irreducible representation $\boldsymbol{N}$ from the appropriate symmetrized product of the defining representations, denoted by powers in parentheses, as in
\eqn{\hat{\mathcal{P}}^{\boldsymbol{N}}\propto\prod_{i=1}^{r-1}(\hat{\mathcal{P}}^{\boldsymbol{e}_i})^{(N_i)}(\hat{\mathcal{P}}^{\boldsymbol{2e}_r})^{(\lfloor N_r/2\rfloor)}(\hat{\mathcal{P}}^{\boldsymbol{e}_r})^{N_r-2\lfloor N_r/2\rfloor}-\text{smaller irreducible representations}.}[EqP]
Here the smaller irreducible representations can be divided into two groups: those representations that are not explicit traces and those that are.  While the latter are trivial to remove via the tracelessness condition, the former are not and can instead be eliminated with the aid of the tensor product decomposition and the projection property.  We point out that here the smaller irreducible representations are not directly subtracted in \eqref{EqP}, as dictated by the completeness relation.  Rather, they are represented by other contributions which encode the proper symmetry properties of the hatted projection operators.

To elucidate this point, we proceed to consider an example.  One of the simplest irreducible representations with mixed symmetry is $\boldsymbol{e}_1+\boldsymbol{e}_2$.  The appropriate form of the hatted projection operator is
\eqn{(\hat{\mathcal{P}}^{\boldsymbol{e}_1+\boldsymbol{e}_2})_{\nu_2\nu_1\mu}^{\phantom{\nu_2\nu_1\mu}\mu'\nu_1'\nu_2'}=\alpha\delta_{\mu}^{\phantom{\mu}\mu'}\delta_{[\nu_1}^{\phantom{[\nu_1}\nu_1'}\delta_{\nu_2]}^{\phantom{\nu_2]}\nu_2'}-\text{smaller irreducible representations},}[EqPe1e2]
according to \eqref{EqPodd}, \eqref{EqPeven}, and \eqref{EqP}.  Here, $\alpha$ is a constant that will be computed later.  We can determine the smaller irreducible representations for $\boldsymbol{e}_1+\boldsymbol{e}_2$ from the tensor product decomposition $\boldsymbol{e}_1\otimes\boldsymbol{e}_2=(\boldsymbol{e}_1+\boldsymbol{e}_2)\oplus\boldsymbol{e}_3\oplus\boldsymbol{e}_1$ where, for notational simplicity, we assume that the rank of the Lorentz group is $r>3$.  It is clear from counting the number of free indices on each smaller irreducible representation, that $\boldsymbol{e}_3$ is not a trace while $\boldsymbol{e}_1$ is.  Thus, $\boldsymbol{e}_1$ is easily subtracted via the tracelessness condition.  Meanwhile, to remove $\boldsymbol{e}_3$, the projection property $\hat{\mathcal{P}}^{\boldsymbol{e}_3}\cdot\hat{\mathcal{P}}^{\boldsymbol{e}_1+\boldsymbol{e}_2}=0$ can be invoked.  But first, it is necessary to determine the form of the contributions of the smaller irreducible representations.  It is clear that we can construct two independent terms, starting from the base term in \eqref{EqPe1e2}, which are
\eqn{\delta_{\mu}^{\phantom{\mu}[\nu_1'}\delta_{[\nu_1}^{\phantom{[\nu_1}\nu_2']}\delta_{\nu_2]}^{\phantom{\nu_2]}\mu'},\qquad\qquad\delta_{\mu[\nu_1}\delta^{\mu'[\nu_1'}\delta_{\nu_2]}^{\phantom{\nu_2]}\nu_2']}.}
These are antisymmetric over the $\nu$ (and $\nu'$) indices, as expected for the hatted projection operator \eqref{EqPe1e2}.  Therefore, the projection operator must be a linear combination of these terms 
\eqn{(\hat{\mathcal{P}}^{\boldsymbol{e}_1+\boldsymbol{e}_2})_{\nu_2\nu_1\mu}^{\phantom{\nu_2\nu_1\mu}\mu'\nu_1'\nu_2'}=\alpha\,\delta_{\mu}^{\phantom{\mu}\mu'}\delta_{[\nu_1}^{\phantom{[\nu_1}\nu_1'}\delta_{\nu_2]}^{\phantom{\nu_2]}\nu_2'}+\beta\,\delta_{\mu}^{\phantom{\mu}[\nu_1'}\delta_{[\nu_1}^{\phantom{[\nu_1}\nu_2']}\delta_{\nu_2]}^{\phantom{\nu_2]}\mu'}+\gamma\,\delta_{\mu[\nu_1}\delta^{\mu'[\nu_1'}\delta_{\nu_2]}^{\phantom{\nu_2]}\nu_2']},}
where $\beta$ and $\gamma$ are two constants that will be fixed shortly.  The two new terms correspond to the contributions due to $\boldsymbol{e}_3$ and $\boldsymbol{e}_1$, respectively.  To see this, we invoke the projection property $\hat{\mathcal{P}}^{\boldsymbol{e}_3}\cdot\hat{\mathcal{P}}^{\boldsymbol{e}_1+\boldsymbol{e}_2}=0$ to fix $\beta=-\alpha$, showing that the addition of the first new term allows us to subtract $\boldsymbol{e}_3$.  Further, the tracelessness condition $g\cdot\hat{\mathcal{P}}^{\boldsymbol{e}_1+\boldsymbol{e}_2}=\hat{\mathcal{P}}^{\boldsymbol{e}_1+\boldsymbol{e}_2}\cdot g=0$ yields $\gamma=-3\alpha/(d-1)$, demonstrating that the last term enables the removal of $\boldsymbol{e}_1$.  Finally, satisfying the projection property $\hat{\mathcal{P}}^{\boldsymbol{e}_1+\boldsymbol{e}_2}\cdot\hat{\mathcal{P}}^{\boldsymbol{e}_1+\boldsymbol{e}_2}=\hat{\mathcal{P}}^{\boldsymbol{e}_1+\boldsymbol{e}_2}$ requires $\alpha=2/3$, thus resulting in the hatted projection operator
\eqn{(\hat{\mathcal{P}}^{\boldsymbol{e}_1+\boldsymbol{e}_2})_{\nu_2\nu_1\mu}^{\phantom{\nu_2\nu_1\mu}\mu'\nu_1'\nu_2'}=\frac{2}{3}\left(\delta_{\mu}^{\phantom{\mu}\mu'}\delta_{[\nu_1}^{\phantom{[\nu_1}\nu_1'}\delta_{\nu_2]}^{\phantom{\nu_2]}\nu_2'}- \delta_{\mu}^{\phantom{\mu}[\nu_1'}\delta_{[\nu_1}^{\phantom{[\nu_1}\nu_2']}\delta_{\nu_2]}^{\phantom{\nu_2]}\mu'}-\frac{3}{d-1}\delta_{\mu[\nu_1}\delta^{\mu'[\nu_1'}\delta_{\nu_2]}^{\phantom{\nu_2]}\nu_2']}\right).}[EqP12]
Obviously, for more complicated irreducible representations it is necessary to construct the hatted projectors for increasingly larger numbers of smaller irreducible representations that are not traces in order to implement the projection property.  This is not always efficient.  Nevertheless, by considering all possible permutations of indices in the base term with the same symmetry properties, it is relatively straightforward to determine all relevant contributions appearing in the hatted projection operators, up to various constants. The latter can then be fixed, in principle, by invoking the projection property and the tracelessness condition.

An alternative approach to constructing the $\hat{\mathcal{P}}^{\boldsymbol{e}_1+\boldsymbol{e}_2}$ projector involves using the completeness relation, as mentioned above.  Up to traces, the only three-index tensors that are antisymmetric in two indices are $\boldsymbol{e}_1+\boldsymbol{e}_2$ and $\boldsymbol{e}_3$.  We can thus derive $\hat{\mathcal{P}}^{\boldsymbol{e}_1+\boldsymbol{e}_2}$ by subtracting the projector to the $\boldsymbol{e}_3$ representation
\eqn{(\hat{\mathcal{P}}^{\boldsymbol{e}_3})_{\nu_2\nu_1\mu}^{\phantom{\nu_2\nu_1\mu}\mu'\nu_1'\nu_2'}=\delta_{[\mu}^{\phantom{[\mu}[\mu'} \delta_{\phantom{[}\!\!\nu_1}^{\phantom{\nu_1}\nu_1'} \delta_{\nu_2]}^{\phantom{\nu_2]}\nu_2']},}
from the identity operator in the subspace of interest and then, as before, discarding the traces.  One obtains
\eqn{(\hat{\mathcal{P}}^{\boldsymbol{e}_1+\boldsymbol{e}_2})_{\nu_2\nu_1\mu}^{\phantom{\nu_2\nu_1\mu}\mu'\nu_1'\nu_2'}=\delta_{\mu}^{\phantom{\mu}\mu'}\delta_{[\nu_1}^{\phantom{[\nu_1}\nu_1'}\delta_{\nu_2]}^{\phantom{\nu_2]}\nu_2'}-\delta_{[\mu}^{\phantom{[\mu}[\mu'}\delta_{\phantom{[}\!\!\nu_1}^{\phantom{[\nu_1}\nu_1'}\delta_{\nu_2]}^{\phantom{\nu_2]}\nu_2']}-\frac{2}{d-1}\delta_{\mu[\nu_1}\delta^{\mu'[\nu_1'}\delta_{\nu_2]}^{\phantom{\nu_2]}\nu_2']}.}[EqP12-alt]
It turns out that the projectors in \eqref{EqP12} and \eqref{EqP12-alt} are identical, as expected.  Note that the overall normalization in \eqref{EqP12-alt} is simply $1$.  This is guaranteed by completeness, because we had arrived at the final form by subtracting other projectors from the identity.  The second term in the equation above is $\hat{\mathcal{P}}^{\boldsymbol{e}_3}$, while $\frac{2}{d-1}\delta_{\mu[\nu_1}\delta^{\mu'[\nu_1'}\delta_{\nu_2]}^{\phantom{\nu_2]}\nu_2']}$ corresponds to the  trace representation $\boldsymbol{e}_1$, which is contained in the product $\boldsymbol{e}_1\otimes\boldsymbol{e}_2$.


\subsection{Half-Projectors}\label{SecHalfP}

The position space half-projectors responsible for the proper behavior of the two-point correlation functions under Lorentz transformations are given by
\eqna{
(\mathcal{T}^{\boldsymbol{N}})_{\alpha_1\cdots\alpha_n}^{\mu_1\cdots\mu_{n_v}\delta}&=\left((\mathcal{T}^{\boldsymbol{e}_1})^{N_1}\cdots(\mathcal{T}^{\boldsymbol{e}_{r-1}})^{N_{r-1}}(\mathcal{T}^{2\boldsymbol{e}_r})^{\lfloor N_r/2\rfloor}(\mathcal{T}^{\boldsymbol{e}_r})^{N_r-2\lfloor N_r/2\rfloor}\right)_{\alpha_1\cdots\alpha_n}^{\mu_1'\cdots\mu_{n_v}'\delta'}\\
&\phantom{=}\qquad\times(\hat{\mathcal{P}}^{\boldsymbol{N}})_{\delta'\mu_{n_v}'\cdots\mu_1'}^{\phantom{\delta'\mu_{n_v}'\cdots\mu_1'}\mu_1\cdots\mu_{n_v}\delta}.
}[EqTPS]
Here $n=2S=2\sum_{i=1}^{r-1}N_i+N_r$ is twice the ``spin'' $S$ of the irreducible representation $\boldsymbol{N}$; $n_v=\sum_{i=1}^{r-1}iN_i+r\lfloor N_r/2\rfloor$ is the number of vector indices of the irreducible representation $\boldsymbol{N}$; and $\delta$ is the spinor index which appears only if $N_r$ is odd (in odd spacetime dimensions).  In \eqref{EqTPS}, the spinor indices $\alpha_1,\cdots,\alpha_n$ match the free indices on the corresponding quasi-primary operator, while the remaining indices $\mu_1,\cdots,\mu_{n_v},\delta$ are dummy indices that are contracted.

Moreover, in \eqref{EqTPS} the position space half-projectors to the defining representations are given by
\eqn{(\mathcal{T}^{\boldsymbol{e}_{i\neq r}})_{\alpha\beta}^{\mu_1\cdots\mu_i}=\frac{1}{\sqrt{2^ri!}}(\gamma^{\mu_1\cdots\mu_i}C^{-1})_{\alpha\beta},\qquad(\mathcal{T}^{\boldsymbol{e}_r})_{\alpha}^{\beta}=\delta_\alpha^{\phantom{\alpha}\beta},\qquad(\mathcal{T}^{2\boldsymbol{e}_r})_{\alpha\beta}^{\mu_1\cdots\mu_r}=\frac{1}{\sqrt{2^rr!}}(\gamma^{\mu_1\cdots\mu_r}C^{-1})_{\alpha\beta},}
where
\eqn{\gamma^{\mu_1\cdots\mu_n}=\frac{1}{n!}\sum_{\sigma\in S_n}(-1)^\sigma\gamma^{\mu_{\sigma(1)}}\cdots\gamma^{\mu_{\sigma(n)}},}
is the totally antisymmetric product of $\gamma$-matrices.

Finally, in \eqref{EqTPS} the hatted projection operator $\hat{\mathcal{P}}^{\boldsymbol{N}}$ contracts with the dummy indices of the half-projector, thus projecting onto the proper irreducible representation $\boldsymbol{N}$.


\subsection{Projectors and Half-Projectors in Embedding Space}

We can easily obtain the embedding space hatted projectors $\hat{\mathcal{P}}_{12}^{\boldsymbol{N}}$ from the corresponding position space quantities [\eqref{EqPodd} and \eqref{EqPeven} for the defining representations, or for any other representation, \textit{e.g.} \eqref{EqP12}] by simply making the following substitutions:
\eqn{
\begin{gathered}
g^{\mu\nu}\to\A_{12}^{AB}\equiv g^{AB}-\frac{\eta_1^A\eta_2^B}{\ee{1}{2}{}}-\frac{\eta_1^B\eta_2^A}{\ee{1}{2}{}},\\
\epsilon^{\mu_1\cdots\mu_d}\to\epsilon_{12}^{A_1\cdots A_d}\equiv\frac{1}{\ee{1}{2}{}}\eta_{1A_0'}\epsilon^{A_0'A_1'\cdots A_d'A_{d+1}'}\eta_{2A_{d+1}'}\A_{12A_d'}^{\phantom{12A_d'}A_d}\cdots\A_{12A_1'}^{\phantom{12A_1'}A_1},\\
\gamma^{\mu_1\cdots\mu_n}\to\Gamma_{12}^{A_1\cdots A_n}\equiv\Gamma^{A_1'\cdots A_n'}\A_{12A_n'}^{\phantom{12A_n'}A_n}\cdots\A_{12A_1'}^{\phantom{12A_1'}A_1}\qquad\forall\,n\in\{0,\ldots,r\},
\end{gathered}
}[EqSubs]
which have the necessary properties (trace, number of vector indices, etc.) to ensure proper contractions with the corresponding irreducible representations in position space.

In embedding space, the corresponding half-projectors are given by
\begingroup\makeatletter\def\f@size{10}\check@mathfonts\def\maketag@@@#1{\hbox{\m@th\large\normalfont#1}}%
\eqna{
(\mathcal{T}_{ij}^{\boldsymbol{N}}\Gamma)&\equiv\left(\left(\frac{\sqrt{2}}{\ee{i}{j}{\frac{1}{2}}}\mathcal{T}^{\boldsymbol{e}_2}\eta_i\A_{ij}\right)^{N_1}\cdots\left(\frac{\sqrt{r}}{\ee{i}{j}{\frac{1}{2}}}\mathcal{T}^{\boldsymbol{e}_{r_E-1}}\eta_i\A_{ij}\cdots\A_{ij}\right)^{N_{r-1}}\right.\\
&\phantom{=}\qquad\times\left.\left(\frac{\sqrt{r+1}}{\ee{i}{j}{\frac{1}{2}}}\mathcal{T}^{2\boldsymbol{e}_{r_E}}\eta_i\A_{ij}\cdots\A_{ij}\right)^{\lfloor N_r/2\rfloor}\left(\frac{1}{\sqrt{2}\ee{i}{j}{}}\mathcal{T}^{\boldsymbol{e}_{r_E}}\eta_i\cdot\Gamma\eta_j\cdot\Gamma\right)^{N_r-2\lfloor N_r/2\rfloor}\right)\cdot\hat{\mathcal{P}}_{ij}^{\boldsymbol{N}},
}[EqTES]
\endgroup
where
\eqn{(\mathcal{T}^{\boldsymbol{e}_{n+1}}\eta_i\A_{ij}\cdots\A_{ij})_{ab}^{A_1\cdots A_n}\equiv(\mathcal{T}^{\boldsymbol{e}_{n+1}})_{ab}^{A_0'\cdots A_n'}\A_{ijA_n'}^{\phantom{ijA_n'}A_n}\cdots\A_{ijA_1'}^{\phantom{ijA_1'}A_1}\eta_{iA_0'}.}
Here the definition of the embedding space half-projectors to the defining representations is the direct analog of the position space definition with the substitutions \eqref{EqSubs} for the projectors and the rank of the Lorentz group $r\to r_E=r+1$, as expected.


\section{Two-Point Correlation Functions in Embedding Space}\label{Sec2ptES}

This section examines two-point correlation functions in the embedding space.  The most general two-point correlation functions of operators in generic Lorentz representations are explicitly given.  All results are presented for the case of odd spacetime dimensions, with the even-dimensional results being a straightforward generalization.

Conformal invariance uniquely specifies the form of the two-point function, up to an overall normalization matrix with indices in the space of the quasi-primary operators, which we refer to as the OPE coefficient matrix.  There is at most one physically allowed two-point structure.  This is transparent from the OPE, which encodes the algebraic structure of the theory.  In fact, from the OPE formalism in \cite{Fortin:2019fvx,Fortin:2019dnq} and as shown below [see also \eqref{EqPC}], non-vanishing two-point structures only exist between quasi-primary operators in irreducible representations $\boldsymbol{N}=\{N_1,\ldots,N_r\}$ and their contragredient-reflected representations $\boldsymbol{N}^{CR}$
\eqna{
\text{d odd:}&\qquad\boldsymbol{N}^{CR}=\{N_1,\ldots,N_r\}=\boldsymbol{N},\\
\text{d even:}&\qquad\boldsymbol{N}^{CR}=\begin{cases}\{N_1,\ldots,N_r\}=\boldsymbol{N},&\text{if $r$ is odd}\\\{N_1,\ldots,N_{r-2},N_r,N_{r-1}\},&\text{if $r$ is even}\end{cases}.
}

Indeed, in arbitrary signature, two-point correlation functions are non-vanishing for representations that are contragredient-reflected with respect to each other.  It is straightforward to see that unless this is true, the proper contraction of the indices is impossible, and the correlator vanishes identically.  In this paper, we restrict attention to CFTs in Lorentzian signature.  In this case, the contragredient-reflected representation $\boldsymbol{N}^{CR}$ is the same as the complex conjugate representation, $\boldsymbol{N}^{CR} =\boldsymbol{N}^C$.  Expressing all two-point functions in Lorentzian signature is convenient for understanding the unitarity conditions, which can be determined by considering two-point correlators between quasi-primary operators and their conjugates.

It is sufficient to include only independent quasi-primary operators for a complete analysis.  On the one hand, we can achieve this by considering all quasi-primary operators and their conjugates; on the other, this can be attained by reducing (almost) all bosonic quasi-primary operators to their real components, thus effectively eliminating the conjugate bosonic quasi-primary operators.  However, the bosonic quasi-primaries in general (anti-)self-dual representations comprise exceptions to this statement if they are not self-conjugate.  Moreover, for fermionic quasi-primaries, such a reduction is possible only for cases when the Majorana condition can be imposed, which corresponds to spacetime dimensions $d=1,2,3\text{ mod }8$ in Lorentzian signature.  Otherwise, conjugate fermionic quasi-primaries are linearly independent and therefore must be included in the OPE.  In view of these observations, in the following, quasi-primary operators and their conjugates are included as long as they are linearly independent.

The two-point correlation functions for quasi-primary operators $\Op{i}{}$ and $\Op{j}{}$ introduced in \eqref{Eq2ptOPE} can be simplified further
\eqn{\Vev{\Op{i}{1}\Op{j}{2}}=(\mathcal{T}_{12}^{\boldsymbol{N}}\Gamma)(\mathcal{T}_{21}^{\boldsymbol{N}^C}\Gamma)\cdot\frac{\lambda_{\boldsymbol{N}}\cOPE{}{i}{j}{\1}\hat{\mathcal{P}}_{12}^{\boldsymbol{N}}}{\ee{1}{2}{\tau}}=(\mathcal{T}_{12}^{\boldsymbol{N}}\Gamma)\cdot(\mathcal{T}_{21}^{\boldsymbol{N}^C}\Gamma)\frac{\lambda_{\boldsymbol{N}}\cOPE{}{i}{j}{\1}}{\ee{1}{2}{\tau}},}[Eq2ptES]
where the hatted projection operator was absorbed into the half-projector, as is evident from \eqref{EqTES}.  Meanwhile, $\lambda_{\boldsymbol{N}}$ is a normalization constant and $\cOPE{}{i}{j}{\1}$ is a matrix of OPE coefficients.

Altogether, the quantity in the numerator of \eqref{Eq2ptES} can be regarded as a group-theoretic part, which constitutes an intertwiner between the representation and its conjugate, serving to effectively join the two representations.  The familiar scalar-like piece $\ee{1}{2}{-\tau}$ is obtained in the standard fashion by seeking the most general function of two points that is Lorentz invariant and homogeneous under scaling transformations.  By construction, this form corresponds to the invariants in $\boldsymbol{N}_i\otimes\boldsymbol{N}_j$, which exist if and only if the Lorentz irreducible representations satisfy $\boldsymbol{N}_i=\boldsymbol{N}_j^C\equiv\boldsymbol{N}$, and the conformal dimensions match $\Delta_i=\Delta_j\equiv\Delta$.  The twist $\tau$ is given by $\tau=\Delta-S$, with $\Delta$ and $S$ denoting the conformal dimension and ``spin'' of the quasi-primary operators, respectively.

The normalization constant $\lambda_{\boldsymbol{N}}$ comes from the tensor structure
\eqn{(\tOPE{}{i}{j}{\1}{1}{2})_{\{aA\}\{bB\}}=\lambda_{\boldsymbol{N}}(\hat{\mathcal{P}}_{12}^{\boldsymbol{N}})_{\{aA\}}^{\phantom{\{aA\}}\{B'b'\}}[(C_\Gamma^{-1})_{b'b}]^{2\xi}(g_{B'B})^{n_v},}
where $(g_{B'B})^{n_v}\equiv g_{B_{n_v}'B_{n_v}}\cdots g_{B_1'B_1}$.  Here $n_v$ denotes the number of vector indices for the Lorentz irreducible representation $\boldsymbol{N}$.  Further, $\xi=S-\lfloor S\rfloor$ vanishes for bosonic operators, while for fermionic operators $\xi=1/2$.  The structure $(\tOPE{}{i}{j}{\1}{1}{2})_{\{aA\}\{bB\}}$ is defined by contracting the projection operator $\hat{\mathcal{P}}_{12}$, with the $g_{A'A}$ metric lowering vector indices, and the $(C_\Gamma^{-1})_{a'a}$ acting as the corresponding metric for spinor indices (see Section 3 in \cite{Fortin:2019dnq} for the conventions on $\Gamma$ matrices).

We choose the normalization constant $\lambda_{\boldsymbol{N}}$ such that the scalar inner product
\eqna{
\tOPE{}{i}{j}{\1}{1}{2}\cdot\tOPE{}{i}{j}{\1*}{2}{1}&\equiv(\tOPE{}{i}{j}{\1}{1}{2})_{\{aA\}\{bB\}}(B_\Gamma^{-1}\tOPE{}{i}{j}{\1*}{2}{1}C_\Gamma^*B_\Gamma C_\Gamma^{-1})_{\{a'A'\}\{b'B'\}}(g^{AA'})^{n_v}[(C_\Gamma)^{aa'}]^{2\xi}(g^{BB'})^{n_v}[(C_\Gamma)^{bb'}]^{2\xi}\\
&=|\lambda_{\boldsymbol{N}}|^2(\hat{\mathcal{P}}_{12}^{\boldsymbol{N}})_{\{aA\}}^{\phantom{\{aA\}}\{Bb\}}(g^{AA'})^{n_v}[(C_\Gamma)^{aa'}]^{2\xi}(B_\Gamma^{-1}\hat{\mathcal{P}}_{21}^{\boldsymbol{N}*}B_\Gamma)_{\{a'A'\}}^{\phantom{\{a'A'\}}\{B'b'\}}[(C_\Gamma^{-1})_{b'b}]^{2\xi}(g_{B'B})^{n_v}\\
&=|\lambda_{\boldsymbol{N}}|^2(\hat{\mathcal{P}}_{12}^{\boldsymbol{N}})_{\{aA\}}^{\phantom{\{aA\}}\{Aa\}}=1
}[EqlambdaN]
is normalized, although its exact value is inconsequential in the following.\footnote{Note however that the normalization constants differ in embedding and position spaces. In embedding space the trace of the identity matrix over spinor indices is twice that of the trace in position space.}  

Note that this tensor structure inner product is different from the one introduced in \cite{Fortin:2019fvx,Fortin:2019dnq}, because the signature here is set to Lorentzian.  This definition of the inner product is thus an artifact of the Lorentzian signature.  This observation also explains why this particular combination of $C_\Gamma$ and $B_\Gamma$ matrices is used in \eqref{EqlambdaN}.  Specifically, the presence of the $B_\Gamma$ matrix in the inner product stems from the definition of the conjugate operators (see Section \ref{SecUnitarity}).

The above identity \eqref{EqlambdaN}, which is shown with the aid of the relation\footnote{It is straightforward to prove the identities $B_\Gamma^{-1}\hat{\mathcal{P}}_{12}^{\boldsymbol{N}*}B_\Gamma=\hat{\mathcal{P}}_{12}^{\boldsymbol{N}^C}$ and \eqref{EqPC} for defining irreducible representations.  By extension, since general irreducible representations are built from the proper (anti-)symmetrization and traces of the defining irreducible representations, these two identities in fact hold for all irreducible representations.} $B_\Gamma^{-1}\hat{\mathcal{P}}_{21}^{\boldsymbol{N}*}B_\Gamma=\hat{\mathcal{P}}_{21}^{\boldsymbol{N}^C}$ and \eqref{EqPC} below, implies that $|\lambda_{\boldsymbol{N}}|^2=[(\hat{\mathcal{P}}_{12}^{\boldsymbol{N}})_{\{aA\}}^{\phantom{\{aA\}}\{Aa\}}]^{-1}$.  Here, the phase on the normalization constant can be chosen such that $\lambda_{\boldsymbol{N}^C}=\lambda_{\boldsymbol{N}}\in\mathbb{R}^+$ without loss of generality.

Upon explicitly exposing all dummy vector and spinor indices, we find the following form for the two-point correlation functions \eqref{Eq2ptES}:
\eqn{\Vev{\Op{i}{1}\Op{j}{2}}=(\mathcal{T}_{12}^{\boldsymbol{N}}\Gamma)^{\{Aa\}}(\mathcal{T}_{21}^{\boldsymbol{N}^C}\Gamma)^{\{Bb\}}[(C_\Gamma^{-1})_{ab}]^{2\xi}(g_{AB})^{n_v}\frac{\lambda_{\boldsymbol{N}}\cOPE{}{i}{j}{\1}}{\ee{1}{2}{\tau}}.}[Eq2ptESDummy]
%


\section{Two-Point Correlation Functions in Position Space}\label{Sec2ptPS}

In this section, we compute two-point correlation functions in position space from the embedding space results in the previous section.  The computations are shown in order of increasing complexity, from the simplest irreducible representations of the Lorentz group to more general ones.

The most important ingredients here are the definitions of the half-projectors $\mathcal{T}_{ij}^{\boldsymbol{N}}\Gamma$ as well as the particular uplift to embedding space and the conventions for Lie algebras (see Section \ref{SecHalfP} and previous results in \cite{Fortin:2019fvx,Fortin:2019dnq} for details).  Moreover, the simple relations
\eqn{x^\mu=\frac{\eta^\mu}{-\eta^{d+1}+\eta^{d+2}},\qquad\ee{1}{2}{}=-\frac{1}{2}(-\eta_1^{d+1}+\eta_1^{d+2})(-\eta_2^{d+1}+\eta_2^{d+2})(x_1-x_2)^2,}[EqetaProd]
between the embedding space and position space coordinates, as well as the light-cone condition $\eta^2=0$, are used extensively in the following.


\subsection{Scalar Quasi-Primary Operators}

For scalar quasi-primary operators, \eqref{Eq2ptESDummy} simplifies greatly to
\eqn{\Vev{\Op{i}{1}\Op{j}{2}}=\frac{\lambda_{\boldsymbol{0}}\cOPE{}{i}{j}{\1}}{\ee{1}{2}{\Delta}},}
due to the vanishing spin.  Although it is of no consequence here, the normalization constant is given by $\lambda_{\boldsymbol{0}}=1$, which follows straightforwardly from \eqref{EqlambdaN}.

Since scalar quasi-primaries do not carry spinor indices, projecting the two-point function from embedding space to position space is trivial and corresponds to $\mathcal{O}^{(x)}(x)=(-\eta^{d+1}+\eta^{d+2})^\Delta\Op{}{}$.  Using \eqref{EqetaProd} and the light-cone condition, we find that the two-point function for scalar quasi-primary operators is simply given by
\eqn{\Vev{\mathcal{O}_i^{(x)}(x_1)\mathcal{O}_j^{(x)}(x_2)}=\frac{\lambda_{\boldsymbol{0}}\cOPE{}{i}{j}{\1}}{\left[-\frac{1}{2}(x_1-x_2)^2\right]^\Delta},}[Eq2ptScalar]
which has exactly the expected form.  Obviously, at this point nothing special has occurred.  However, it will become apparent from more complicated examples which follow that all irreducible representations are treated in a unified fashion in this formalism.


\subsection{Spinor Quasi-Primary Operators}

We next consider the defining spinor representations.  Since these differ according to the spacetime dimension, we treat the odd- and even-dimensional cases separately.


\subsubsection{Odd Dimensions: $p=1$, $q=d-1$ and $d=p+q=2r+1$}

In odd spacetime dimensions, there is only one irreducible spinor representation.  From the general form in \eqref{Eq2ptESDummy}, the two-point correlation functions are simply given by
\eqn{\Vev{\Op{i}{1}\Op{j}{2}}=(\mathcal{T}_{12}^{\boldsymbol{e}_r}\Gamma)^a(\mathcal{T}_{21}^{\boldsymbol{e}_r^C}\Gamma)^b(\hat{\mathcal{P}}_{12}^{\boldsymbol{e}_r})_a^{\phantom{a}b'}(C_\Gamma^{-1})_{b'b}\frac{\lambda_{\boldsymbol{e}_r}\cOPE{}{i}{j}{\1}}{\ee{1}{2}{\tau}}.}
Here the hatted projection operator is $(\hat{\mathcal{P}}_{12}^{\boldsymbol{e}_r})_a^{\phantom{a}b'}=\delta_a^{\phantom{a}b'}$, and hence two-point correlation functions for spinor quasi-primary operators with embedding space spinor indices take the form
\eqna{
\Vev{\Op{ia}{1}\Op{jb}{2}}&=\frac{1}{2\ee{1}{2}{2}}(\eta_1\cdot\Gamma\eta_2\cdot\Gamma)_a^{\phantom{a}a'}(\eta_2\cdot\Gamma\eta_1\cdot\Gamma)_b^{\phantom{b}b'}(C_\Gamma^{-1})_{a'b'}\frac{\lambda_{\boldsymbol{e}_r}\cOPE{}{i}{j}{\1}}{\ee{1}{2}{\Delta-1/2}}\\
&=(\eta_1\cdot\Gamma\eta_2\cdot\Gamma C_\Gamma^{-1})_{ab}\frac{\lambda_{\boldsymbol{e}_r}\cOPE{}{i}{j}{\1}}{\ee{1}{2}{\Delta+1/2}},
}
where the embedding space matrices have been properly simplified and $\lambda_{\boldsymbol{e}_r}=\sqrt{1/2^{r+1}}$ from \eqref{EqlambdaN}.  It is of interest to point out here that the tensor structure, which is proportional to the hatted projection operator, contracts the two defining spinor representations into a singlet in embedding space, which contrasts with the situation in position space.

We now project this expression to position space.  This implies keeping only the first half of the embedding space spinor indices for each of the two quasi-primary operators and multiplying by the proper homogeneity factor, which corresponds to $\mathcal{O}^{(x)}(x)=(-\eta^{d+1}+\eta^{d+2})^{\Delta-1/2}\Op{+}{}$.  Hence, in the product $\Gamma^A\Gamma^BC_\Gamma^{-1}$ only the first diagonal block element $M_{11}$ in the block matrix representation $\left(\begin{array}{cc}M_{11}&M_{12}\\M_{21}&M_{22}\end{array}\right)$ is relevant.  All the other elements project to zero and therefore do not contribute.  Since from their definitions (see \cite{Fortin:2019fvx,Fortin:2019dnq}) $\Gamma^\mu$ are block diagonal while $\Gamma^{d+1}$, $\Gamma^{d+2}$ and $C_\Gamma$ are block off-diagonal, projection to position space constrains $A$ to be $\mu$ and $B$ to be $d+1, d+2$, or vice versa.

It emerges that the only relevant part of $(\eta_1\cdot\Gamma\eta_2\cdot\Gamma C_\Gamma^{-1})_{ab}$ in position space is just
\eqn{[\eta_{1\mu}\Gamma^\mu(\eta_{2,d+1}\Gamma^{d+1}+\eta_{2,d+2}\Gamma^{d+2})C_\Gamma^{-1}+(\eta_{1,d+1}\Gamma^{d+1}+\eta_{1,d+2}\Gamma^{d+2})\eta_{2\mu}\Gamma^\mu C_\Gamma^{-1}]_{\alpha\beta}.}
Now, the explicit form of the matrices in embedding space in terms of their position space counterparts gives
\eqn{\alpha[(-\eta_2^{d+1}+\eta_2^{d+2})\eta_{1\mu}-(-\eta_1^{d+1}+\eta_1^{d+2})\eta_{2\mu}](\gamma^\mu C^{-1})_{\alpha\beta}.
}
Converting between the position space and embedding space coordinates, the two-point function is
\eqn{\Vev{\mathcal{O}_{i\alpha}^{(x)}(x_1)\mathcal{O}_{j\beta}^{(x)}(x_2)}=\alpha(x_1-x_2)_\mu(\gamma^\mu C^{-1})_{\alpha\beta}\frac{\lambda_{\boldsymbol{e}_r}\cOPE{}{i}{j}{\1}}{\left[-\frac{1}{2}(x_1-x_2)^2\right]^{\Delta+1/2}},}[Eq2ptSpinorOdd]
where \eqref{EqetaProd} and the light-cone condition have been used.  Once again, we see that this is the expected form from conformal covariance.


\subsubsection{Even Dimensions: $p=1$, $q=d-1$ and $d=p+q=2r$}

We next turn to the case of even spacetime dimensions $d=p+q=2r$.  There exist two inequivalent irreducible spinor representations, namely $\boldsymbol{e}_{r-1}$ and $\boldsymbol{e}_r$, in contrast to the odd-dimensional case.  As explained above, their behavior under charge conjugation depends on the rank and signature of the Lorentz group of interest.  Here we consider the case of $SO(1,d-1)$ so that the signature is fixed, since $q$ is always odd.  We are therefore left with only two separate cases to consider, namely $r$ even and $r$ odd.

In position space, Lorentz covariance constrains the non-vanishing two-point correlation functions to quasi-primary operators in conjugate representations.  Since $q$ is odd, this implies that both quasi-primary operators are in different (the same) irreducible spinor representations for $r$ even (odd).  To ensure proper contraction of the embedding space spinor indices, we note that we must take into account $r_E=r+1$ and $q_E = q+1$ in the embedding space matrices.  This fact effectively changes the parity of both parameters in the embedding space, thereby properly restricting all embedding space spinor index contractions.

With this in mind, we observe that in even spacetime dimensions for $r$ even, the general two-point function \eqref{Eq2ptESDummy} form reduces to
\eqna{
\left.\Vev{\Op{i}{1}\Op{j}{2}}\right|_\text{$r$ even}&=(\mathcal{T}_{12}^{\boldsymbol{e}_{r-1}}\Gamma)^a(\mathcal{T}_{21}^{\boldsymbol{e}_{r-1}^C}\Gamma)^{\tilde{b}}(\hat{\mathcal{P}}_{12}^{\boldsymbol{e}_{r-1}})_a^{\phantom{a}b'}(C_\Gamma^{-1})_{b'\tilde{b}}\frac{\lambda_{\boldsymbol{e}_{r-1}}\cOPE{}{i}{j}{\1}}{\ee{1}{2}{\tau}}\\
&=\frac{1}{2\ee{1}{2}{2}}(\eta_1\cdot\Gamma\eta_2\cdot\tilde{\Gamma})^a(\eta_2\cdot\tilde{\Gamma}\eta_1\cdot\Gamma)^{\tilde{b}}(C_\Gamma^{-1})_{a\tilde{b}}\frac{\lambda_{\boldsymbol{e}_{r-1}}\cOPE{}{i}{j}{\1}}{\ee{1}{2}{\Delta-1/2}},
}
since the hatted projection operator is trivial, $(\hat{\mathcal{P}}_{12}^{\boldsymbol{e}_{r-1}})_a^{\phantom{a}b'}=\delta_a^{\phantom{a}b'}$.  Upon reintroducing the quasi-primary operator spinor indices and commuting the matrices through, we obtain
\eqn{\left.\Vev{\Op{ia}{1}\Op{j\tilde{b}}{2}}\right|_\text{$r$ even}=(\eta_1\cdot\Gamma\eta_2\cdot\tilde{\Gamma}C_\Gamma^{-1})_{a\tilde{b}}\frac{\lambda_{\boldsymbol{e}_{r-1}}\cOPE{}{i}{j}{\1}}{\ee{1}{2}{\Delta+1/2}},}
with $\lambda_{\boldsymbol{e}_{r-1}}=\sqrt{1/2^r}$ from \eqref{EqlambdaN}.

Employing identical reasoning for the case of $r$ odd leads to the two-point correlation functions
\eqna{
\left.\Vev{\Op{ia}{1}\Op{jb}{2}}\right|_\text{$r$ odd}&=(\mathcal{T}_{12}^{\boldsymbol{e}_{r-1}}\Gamma)_a^{\phantom{a}a'}(\mathcal{T}_{21}^{\boldsymbol{e}_{r-1}^C}\Gamma)_b^{\phantom{b}b'}(\hat{\mathcal{P}}_{12}^{\boldsymbol{e}_{r-1}})_{a'}^{\phantom{a'}b''}(C_\Gamma^{-1})_{b''b'}\frac{\lambda_{\boldsymbol{e}_{r-1}}\cOPE{}{i}{j}{\1}}{\ee{1}{2}{\tau}}\\
&=(\eta_1\cdot\Gamma\eta_2\cdot\tilde{\Gamma}C_\Gamma^{-1})_{ab}\frac{\lambda_{\boldsymbol{e}_{r-1}}\cOPE{}{i}{j}{\1}}{\ee{1}{2}{\Delta+1/2}},\\
\left.\Vev{\Op{i\tilde{a}}{1}\Op{j\tilde{b}}{2}}\right|_\text{$r$ odd}&=(\mathcal{T}_{12}^{\boldsymbol{e}_r}\Gamma)_{\tilde{a}}^{\phantom{\tilde{a}}\tilde{a}'}(\mathcal{T}_{21}^{\boldsymbol{e}_r^C}\Gamma)_{\tilde{b}}^{\phantom{\tilde{b}}\tilde{b}'}(\hat{\mathcal{P}}_{12}^{\boldsymbol{e}_r})_{\tilde{a}'}^{\phantom{\tilde{a}'}\tilde{b}''}(\tilde{C}_\Gamma^{-1})_{\tilde{b}''\tilde{b}'}\frac{\lambda_{\boldsymbol{e}_r}\cOPE{}{i}{j}{\1}}{\ee{1}{2}{\tau}}\\
&=(\eta_1\cdot\tilde{\Gamma}\eta_2\cdot\Gamma\tilde{C}_\Gamma^{-1})_{\tilde{a}\tilde{b}}\frac{\lambda_{\boldsymbol{e}_r}\cOPE{}{i}{j}{\1}}{\ee{1}{2}{\Delta+1/2}},
}
with $\lambda_{\boldsymbol{e}_{r-1}}=\lambda_{\boldsymbol{e}_r}=\sqrt{1/2^r}$ from \eqref{EqlambdaN}.

We subsequently project these expressions to position space, proceeding in the same fashion as for the odd-dimensional case, which gives
\eqna{
\left.\Vev{\mathcal{O}_{i\alpha}^{(x)}(x_1)\mathcal{O}_{j\tilde{\beta}}^{(x)}(x_2)}\right|_\text{$r$ even}&=\alpha(x_1-x_2)_\mu(\gamma^\mu\tilde{C}^{-1})_{\alpha\tilde{\beta}}\frac{\lambda_{\boldsymbol{e}_{r-1}}\cOPE{}{i}{j}{\1}}{\left[-\frac{1}{2}(x_1-x_2)^2\right]^{\Delta+1/2}},\\
\left.\Vev{\mathcal{O}_{i\alpha}^{(x)}(x_1)\mathcal{O}_{j\beta}^{(x)}(x_2)}\right|_\text{$r$ odd}&=\alpha(x_1-x_2)_\mu(\gamma^\mu\tilde{C}^{-1})_{\alpha\beta}\frac{\lambda_{\boldsymbol{e}_{r-1}}\cOPE{}{i}{j}{\1}}{\left[-\frac{1}{2}(x_1-x_2)^2\right]^{\Delta+1/2}},\\
\left.\Vev{\mathcal{O}_{i\tilde{\alpha}}^{(x)}(x_1)\mathcal{O}_{j\tilde{\beta}}^{(x)}(x_2)}\right|_\text{$r$ odd}&=\alpha(x_1-x_2)_\mu(\tilde{\gamma}^\mu C^{-1})_{\tilde{\alpha}\tilde{\beta}}\frac{\lambda_{\boldsymbol{e}_r}\cOPE{}{i}{j}{\1}}{\left[-\frac{1}{2}(x_1-x_2)^2\right]^{\Delta+1/2}},
}[Eq2ptSpinorEven]
exactly as expected from conformal covariance.


\subsection{Antisymmetric Quasi-Primary Operators}

We now go on to consider the remaining defining representations, the $n$-index antisymmetric tensors.  Utilizing \eqref{Eq2ptESDummy}, we see that their two-point correlation functions are given by
\eqn{\Vev{\Op{i}{1}\Op{j}{2}}=(\mathcal{T}_{12}^{\boldsymbol{e}_n}\Gamma)^{A_1\cdots A_n}(\mathcal{T}_{21}^{\boldsymbol{e}_n^C}\Gamma)^{B_1\cdots B_n}(\hat{\mathcal{P}}_{12}^{\boldsymbol{e}_n})_{A_n\cdots A_1}^{\phantom{A_n\cdots A_1}B_1'\cdots B_n'}g_{B_n'B_n}\cdots g_{B_1'B_1}\frac{\lambda_{\boldsymbol{e}_n}\cOPE{}{i}{j}{\1}}{\ee{1}{2}{\tau}},}
where it is understood that the $r$-index antisymmetric representation for odd spacetime dimensions is denoted by $2\boldsymbol{e}_r$, while for even dimensions the $(r-1)$-index, the self-dual $r$-index, and the anti-self-dual $r$-index antisymmetric representations are referred to as $\boldsymbol{e}_{r-1}+\boldsymbol{e}_r$, $2\boldsymbol{e}_{r-1}$, and $2\boldsymbol{e}_r$, respectively.

From \eqref{EqSubs} the hatted projection operator is simply $(\hat{\mathcal{P}}_{12}^{\boldsymbol{e}_n})_{A_n\cdots A_1}^{\phantom{A_n\cdots A_1}B_1'\cdots B_n'}=\A_{12[A_1}^{\phantom{12[A_1}B_1'}\cdots\A_{12A_n]}^{\phantom{12A_n]}B_n'}$, where the $A$-indices (and by proxy the $B'$-indices) are fully antisymmetrized.  This applies to all $n$-index antisymmetric representations except the self-dual and anti-self-dual representations in even spacetime dimensions, for which
\eqna{
(\hat{\mathcal{P}}_{12}^{2\boldsymbol{e}_{r-1}})_{A_r\cdots A_1}^{\phantom{A_r\cdots A_1}B_1'\cdots B_r'}&=\frac{1}{2}\A_{12[A_1}^{\phantom{12[A_1}B_1'}\cdots\A_{12A_r]}^{\phantom{12A_r]}B_r'}+(-1)^r\frac{\mathscr{K}}{2r!}\epsilon_{12A_1\cdots A_r}^{\phantom{12A_1\cdots A_r}B_r'\cdots B_1'},\\
(\hat{\mathcal{P}}_{12}^{2\boldsymbol{e}_r})_{A_r\cdots A_1}^{\phantom{A_r\cdots A_1}B_1'\cdots B_r'}&=\frac{1}{2}\A_{12[A_1}^{\phantom{12[A_1}B_1'}\cdots\A_{12A_r]}^{\phantom{12A_r]}B_r'}-(-1)^r\frac{\mathscr{K}}{2r!}\epsilon_{12A_1\cdots A_r}^{\phantom{12A_1\cdots A_r}B_r'\cdots B_1'}.
}

Since the half-projectors are already fully antisymmetrized in their two sets of dummy indices, the two-point correlation functions assume the form
\eqna{
\Vev{\Op{ia_1a_2}{1}\Op{jb_1b_2}{2}}&=\frac{n+1}{\ee{1}{2}{}}(\mathcal{T}^{\boldsymbol{e}_{n+1}}\eta_1\A_{12}\cdots\A_{12})_{a_1a_2}^{A_1\cdots A_n}(\mathcal{T}^{\boldsymbol{e}_{n+1}^C}\eta_2\A_{12}\cdots\A_{12})_{b_1b_2}^{B_1\cdots B_n}\\
&\phantom{=}\qquad\times g_{A_nB_n}\cdots g_{A_1B_1}\frac{\lambda_{\boldsymbol{e}_n}\cOPE{}{i}{j}{\1}}{\ee{1}{2}{\Delta-1}}\\
&=\frac{1}{2^{\lfloor(d+1)/2\rfloor}n!}(\Gamma^{A_0\cdots A_n}C_\Gamma^{-1})_{a_1a_2}(\Gamma^{B_0\cdots B_n}C_\Gamma^{-1})_{b_1b_2}\\
&\phantom{=}\qquad\times\eta_{1A_0}\eta_{2B_0}\A_{12A_1B_1}\cdots\A_{12A_nB_n}\frac{\lambda_{\boldsymbol{e}_n}\cOPE{}{i}{j}{\1}}{\ee{1}{2}{\Delta}},
}
where for simplicity we take the embedding space spinor indices on the quasi-primary operators to be without tildes (the other cases are similar).  We remark here that $\lambda_{\boldsymbol{e}_n}=\sqrt{n!/(d+1-n)_n}$ where $n=r$ for $2\boldsymbol{e}_r$ in odd spacetime dimensions, and $n=r-1$ for $\boldsymbol{e}_{r-1}+\boldsymbol{e}_r$ in even spacetime dimensions, while $\lambda_{2\boldsymbol{e}_{r-1}}=\lambda_{2\boldsymbol{e}_r}=\sqrt{2r!/(d+1-r)_r}$ for the (anti-)self-dual irreducible representations in even spacetime dimensions.

With the aid of the identity
\eqn{\Gamma^{A_0\cdots A_n}=\Gamma^{A_0}\Gamma^{A_1\cdots A_n}+\sum_{i=1}^n(-1)^ig^{A_0A_i}\Gamma^{A_1\cdots\widehat{A}_i\cdots A_n},}
we can further simplify the embedding space two-point functions to
\eqn{\Vev{\Op{ia_1a_2}{1}\Op{jb_1b_2}{2}}=\frac{1}{2^{\lfloor(d+1)/2\rfloor}n!}(\eta_1\cdot\Gamma\Gamma_{12}^{A_1\cdots A_n}C_\Gamma^{-1})_{a_1a_2}(\eta_2\cdot\Gamma\Gamma_{A_1\cdots A_n}C_\Gamma^{-1})_{b_1b_2}\frac{\lambda_{\boldsymbol{e}_n}\cOPE{}{i}{j}{\1}}{\ee{1}{2}{\Delta}},}
where we have taken advantage of the double-transversality property of the metric $\A_{12}$, \textit{i.e.} $\eta_1\cdot\A_{12}=\eta_2\cdot\A_{12}=0$.

We now project to position space exactly as before by restricting to the first half of all embedding space spinor indices, effectively picking up the top left block in the block matrix representation of $\Gamma^{A_0\cdots A_n}C_\Gamma^{-1}$ and $\Gamma^{B_0\cdots B_n}C_\Gamma^{-1}$.  Since these products are fully antisymmetric, at most one index can be $d+1$, and one $d+2$.  It is evident from the form of the embedding space $\Gamma$ matrices in terms of position space matrices that the only nonzero contributions arise from the products carrying either one $d+1$ index or one $d+2$ index.

Thus, the only relevant contributions of $(\eta_1\cdot\Gamma_{12}^{A_1\cdots A_n}C_\Gamma^{-1})_{a_1a_2}(\eta_2\cdot\Gamma_{A_1\cdots A_n}C_\Gamma^{-1})_{b_1b_2}$ in position space are
\eqna{
&\{-n(n-1)(\eta_{1\mu_1}\Gamma^{\mu_1\mu_2\cdots\mu_nd+1}C_\Gamma^{-1})_{\alpha_1\alpha_2}(\eta_{2\nu_1}\Gamma^{\nu_1\nu_2\cdots\nu_nd+1}C_\Gamma^{-1})_{\beta_1\beta_2}\A_{12\mu_2,d+1}\A_{12,d+1,\nu_2}\\
&\qquad-n(n-1)(\eta_{1\mu_1}\Gamma^{\mu_1\mu_2\cdots\mu_nd+1}C_\Gamma^{-1})_{\alpha_1\alpha_2}(\eta_{2\nu_1}\Gamma^{\nu_1\nu_2\cdots\nu_nd+2}C_\Gamma^{-1})_{\beta_1\beta_2}\A_{12\mu_2,d+2}\A_{12,d+1,\nu_2}\\
&\qquad-n(n-1)(\eta_{1\mu_1}\Gamma^{\mu_1\mu_2\cdots\mu_nd+2}C_\Gamma^{-1})_{\alpha_1\alpha_2}(\eta_{2\nu_1}\Gamma^{\nu_1\nu_2\cdots\nu_nd+1}C_\Gamma^{-1})_{\beta_1\beta_2}\A_{12\mu_2,d+1}\A_{12,d+2,\nu_2}\\
&\qquad-n(n-1)(\eta_{1\mu_1}\Gamma^{\mu_1\mu_2\cdots\mu_nd+2}C_\Gamma^{-1})_{\alpha_1\alpha_2}\\
&\qquad\qquad\times(\eta_{2\nu_1}\Gamma^{\nu_1\nu_2\cdots\nu_nd+2}C_\Gamma^{-1})_{\beta_1\beta_2}\A_{12\mu_2,d+2}\A_{12,d+2,\nu_2}\}\A_{12\mu_3\nu_3}\cdots\A_{12\mu_n\nu_n}\\
&\qquad+\{n(\eta_{1\mu_1}\Gamma^{\mu_1\mu_2\cdots\mu_nd+1}C_\Gamma^{-1})_{\alpha_1\alpha_2}(\eta_{2\nu_1}\Gamma^{\nu_1\nu_2\cdots\nu_nd+1}C_\Gamma^{-1})_{\beta_1\beta_2}\A_{12,d+1,d+1}\\
&\qquad+n(\eta_{1\mu_1}\Gamma^{\mu_1\mu_2\cdots\mu_nd+1}C_\Gamma^{-1})_{\alpha_1\alpha_2}(\eta_{2\nu_1}\Gamma^{\nu_1\nu_2\cdots\nu_nd+2}C_\Gamma^{-1})_{\beta_1\beta_2}\A_{12,d+1,d+2}\\
&\qquad+n(\eta_{1\mu_1}\Gamma^{\mu_1\mu_2\cdots\mu_nd+2}C_\Gamma^{-1})_{\alpha_1\alpha_2}(\eta_{2\nu_1}\Gamma^{\nu_1\nu_2\cdots\nu_nd+1}C_\Gamma^{-1})_{\beta_1\beta_2}\A_{12,d+2,d+1}\\
&\qquad+n(\eta_{1\mu_1}\Gamma^{\mu_1\mu_2\cdots\mu_nd+2}C_\Gamma^{-1})_{\alpha_1\alpha_2}(\eta_{2\nu_1}\Gamma^{\nu_1\nu_2\cdots\nu_nd+2}C_\Gamma^{-1})_{\beta_1\beta_2}\A_{12,d+2,d+2}\\
&\qquad-(-1)^nn(\eta_{1\mu_1}\Gamma^{\mu_1\mu_2\cdots\mu_nd+1}C_\Gamma^{-1})_{\alpha_1\alpha_2}[(\eta_{2,d+1}\Gamma^{d+1}+\eta_{2,d+2}\Gamma^{d+2})^{\nu_1\cdots\nu_n}C_\Gamma^{-1}]_{\beta_1\beta_2}\A_{12,d+1,\nu_1}\\
&\qquad-(-1)^nn(\eta_{1\mu_1}\Gamma^{\mu_1\mu_2\cdots\mu_nd+2}C_\Gamma^{-1})_{\alpha_1\alpha_2}[(\eta_{2,d+1}\Gamma^{d+1}+\eta_{2,d+2}\Gamma^{d+2})^{\nu_1\cdots\nu_n}C_\Gamma^{-1}]_{\beta_1\beta_2}\A_{12,d+2,\nu_1}\\
&\qquad-(-1)^nn[(\eta_{1,d+1}\Gamma^{d+1}+\eta_{1,d+2}\Gamma^{d+2})^{\mu_1\cdots\mu_n}C_\Gamma^{-1}]_{\alpha_1\alpha_2}(\eta_{2\nu_1}\Gamma^{\nu_1\nu_2\cdots\nu_nd+1}C_\Gamma^{-1})_{\beta_1\beta_2}\A_{12\mu_1,d+1}\\
&\qquad-(-1)^nn[(\eta_{1,d+1}\Gamma^{d+2}+\eta_{1,d+2}\Gamma^{d+2})^{\mu_1\cdots\mu_n}C_\Gamma^{-1}]_{\alpha_1\alpha_2}(\eta_{2\nu_1}\Gamma^{\nu_1\nu_2\cdots\nu_nd+1}C_\Gamma^{-1})_{\beta_1\beta_2}\A_{12\mu_1,d+2}\\
&\qquad+[(\eta_{1,d+1}\Gamma^{d+1}+\eta_{1,d+2}\Gamma^{d+2})^{\mu_1\cdots\mu_n}C_\Gamma^{-1}]_{\alpha_1\alpha_2}\\
&\qquad\qquad\times[(\eta_{2,d+1}\Gamma^{d+1}+\eta_{2,d+2}\Gamma^{d+2})^{\nu_1\cdots\nu_n}C_\Gamma^{-1}]_{\beta_1\beta_2}\A_{12\mu_1\nu_1}\}\A_{12\mu_2\nu_2}\cdots\A_{12\mu_n\nu_n}.
}
The factors of $-n(n-1)$, $n$ and $-(-1)^nn$ can be understood from permutations of the $d+1$ or $d+2$ indices to the last position.  Indeed, the factor $-n(n-1)$ represents the number of ways the indices $d+1$ or $d+2$ can occur through the various $\A_{12}$ metrics.  The factor $n$ comes from the number of ways the indices $d+1$ or $d+2$ can occur through the same $\A_{12}$ metric.  Finally, the factor $-(-1)^nn$ originates from the number of ways the index $d+1$ or $d+2$ can occur through a $\A_{12}$ metric.

Extracting the matrices $\Gamma^{d+1}$ or $\Gamma^{d+2}$ as, for example, in $\Gamma^{\mu_1\cdots\mu_nd+1}=\Gamma^{\mu_1\cdots\mu_n}\Gamma^{d+1}$, we find the following form in terms of position space matrices:
\eqna{
&\alpha^2(-\eta_1^{d+1}+\eta_1^{d+2})(-\eta_2^{d+1}+\eta_2^{d+2})\left\{-\frac{4n(n-1)x_{1\mu_1}x_{2\nu_1}(x_1+x_2)_{\mu_2}(x_1+x_2)_{\nu_2}}{(x_1-x_2)^4}\right.\\
&\qquad\left.+\left[\frac{nx_{1\mu_1}x_{2\nu_1}}{(x_1-x_2)^2}-\frac{2nx_{1\mu_1}(x_1+x_2)_{\nu_1}}{(x_1-x_2)^2}-\frac{2n(x_1+x_2)_{\mu_1}x_{2\nu_1}}{(x_1-x_2)^2}+\A_{12\mu_1\nu_1}\right]\A_{12\mu_2\nu_2}\right\}\\
&\qquad\times\A_{12\mu_3\nu_3}\cdots\A_{12\mu_n\nu_n}(\gamma^{\mu_1\cdots\mu_n}C^{-1})_{\alpha_1\alpha_2}(\gamma^{\nu_1\cdots\nu_n}C^{-1})_{\beta_1\beta_2}\\
&=\alpha^2(-\eta_1^{d+1}+\eta_1^{d+2})(-\eta_2^{d+1}+\eta_2^{d+2})\\
&\qquad\times\left[\A_{12\mu_1\nu_1}\A_{12\mu_2\nu_2}+n\mathcal{B}_{12\mu_1\nu_1}\A_{12\mu_2\nu_2}+\frac{n(n-1)}{2}\mathcal{B}_{12\mu_1\nu_1}\mathcal{B}_{12\mu_2\nu_2}\right]\\
&\qquad\times\A_{12\mu_3\nu_3}\cdots\A_{12\mu_n\nu_n}(\gamma^{\mu_1\cdots\mu_n}C^{-1})_{\alpha_1\alpha_2}(\gamma^{\nu_1\cdots\nu_n}C^{-1})_{\beta_1\beta_2}\\
&=\alpha^2(-\eta_1^{d+1}+\eta_1^{d+2})(-\eta_2^{d+1}+\eta_2^{d+2})\\
&\qquad\times I_{\mu_1\nu_1}(x_1-x_2)\cdots I_{\mu_n\nu_n}(x_1-x_2)(\gamma^{\mu_1\cdots\mu_n}C^{-1})_{\alpha_1\alpha_2}(\gamma^{\nu_1\cdots\nu_n}C^{-1})_{\beta_1\beta_2},
}
where we have used the simple relations
\eqn{
\begin{gathered}
\A_{12\mu\nu}=I_{\mu\nu}(x_1-x_2)-\mathcal{B}_{12\mu\nu},\qquad I_{\mu\nu}(x)=g_{\mu\nu}-2\frac{x_\mu x_\nu}{x^2},\qquad\mathcal{B}_{12\mu\nu}=-2\frac{x_{1\mu}x_{1\nu}+x_{2\mu}x_{2\nu}}{(x_1-x_2)^2},\\
\A_{12\mu,d+1}+\A_{12\mu,d+2}=2\frac{(x_1+x_2)_\mu}{(x_1-x_2)^2},\\
\A_{12,d+1,d+1}+2\A_{12,d+1,d+2}+\A_{12,d+2,d+2}=\frac{4}{(x_1-x_2)^2},
\end{gathered}
}
and the antisymmetrization property of the matrices.

We can now straightforwardly obtain the position space two-point functions for $n$-index antisymmetric quasi-primary operators:
\eqna{
\Vev{\mathcal{O}_{i\alpha_1\alpha_2}^{(x)}(x_1)\mathcal{O}_{j\beta_1\beta_2}^{(x)}(x_2)}&=\frac{\alpha^2}{2^{\lfloor(d+1)/2\rfloor}n!}(\gamma^{\mu_1\cdots\mu_n}C^{-1})_{\alpha_1\alpha_2}(\gamma^{\nu_1\cdots\nu_n}C^{-1})_{\beta_1\beta_2}\\
&\phantom{=}\qquad\times I_{\mu_1\nu_1}(x_1-x_2)\cdots I_{\mu_n\nu_n}(x_1-x_2)\frac{\lambda_{\boldsymbol{e}_n}\cOPE{}{i}{j}{\1}}{\left[-\frac{1}{2}(x_1-x_2)^2\right]^\Delta}\\
&=\frac{\alpha^2}{2}(\mathcal{T}^{\boldsymbol{e}_n})_{\alpha_1\alpha_2}^{\mu_1\cdots\mu_n}(\mathcal{T}^{\boldsymbol{e}_n^C})_{\beta_1\beta_2}^{\nu_1\cdots\nu_n}\\
&\phantom{=}\qquad\times I_{\mu_1\nu_1}(x_1-x_2)\cdots I_{\mu_n\nu_n}(x_1-x_2)\frac{\lambda_{\boldsymbol{e}_n}\cOPE{}{i}{j}{\1}}{\left[-\frac{1}{2}(x_1-x_2)^2\right]^\Delta},
}[Eq2ptAntisym]
again in perfect agreement with covariance under the conformal group, as indicated by the appearance of the inversion structure $I_{\mu\nu}(x)$.\footnote{Although inversions are not elements of the conformal group that are connected to the identity, a special conformal transformation can be seen as an inversion followed by a translation and another inversion.}  Note that for (anti-)self-dual representations, the presence of the conjugate is crucial in the position-space result \eqref{Eq2ptAntisym} (after reintroducing the proper tilde or untilde spinor indices required in even dimensions), just as it originally was in the embedding space \eqref{Eq2ptES}.  Indeed, two-point correlation functions of (anti-)self-dual quasi-primaries are non-vanishing provided that the quasi-primary operators are conjugates of one another, which is a straightforward observation in the context of the embedding space [see \eqref{EqPC}].


\subsection{General Quasi-Primary Operators}\label{SecGeneral}

Quasi-primary operators in general irreducible representations of the Lorentz group can be constructed from the defining irreducible representations already discussed.  The most general two-point correlation functions are given by \eqref{Eq2ptESDummy}, and explicit computations reveal that position space two-point correlation functions have the form
\eqna{
\Vev{\mathcal{O}_i^{(x)\boldsymbol{N}}(x_1)\mathcal{O}_j^{(x)\boldsymbol{N}^C}(x_2)}&=\left(\frac{\alpha^2}{2}\right)^{S-\xi}(\mathcal{T}^{\boldsymbol{N}})^{\{\mu\delta\}}(\mathcal{T}^{\boldsymbol{N}^C})^{\{\nu\epsilon\}}(\hat{\mathcal{P}}^{\boldsymbol{N}})_{\{\delta\mu\}}^{\phantom{\{\delta\mu\}}\{\nu'\epsilon'\}}\\
&\phantom{=}\qquad\times[\alpha(x_1-x_2)\cdot(\gamma C^{-1})_{\epsilon'\epsilon}]^{2\xi}[I_{\nu'\nu}(x_1-x_2)]^{n_v}\frac{\lambda_{\boldsymbol{N}}\cOPE{}{i}{j}{\1}}{\left[-\frac{1}{2}(x_1-x_2)^2\right]^{\Delta+\xi}}\\
&=\left(\frac{\alpha^2}{2}\right)^{S-\xi}(\mathcal{T}^{\boldsymbol{N}})^{\{\mu\delta\}}(\mathcal{T}^{\boldsymbol{N}^C})^{\{\nu\epsilon\}}\\
&\phantom{=}\qquad\times[\alpha(x_1-x_2)\cdot(\gamma C^{-1})_{\delta\epsilon}]^{2\xi}[I_{\mu\nu}(x_1-x_2)]^{n_v}\frac{\lambda_{\boldsymbol{N}}\cOPE{}{i}{j}{\1}}{\left[-\frac{1}{2}(x_1-x_2)^2\right]^{\Delta+\xi}}.
}[Eq2ptPS]
This result is a direct analog of the formulas \eqref{Eq2ptScalar}, \eqref{Eq2ptSpinorOdd}, \eqref{Eq2ptSpinorEven} and \eqref{Eq2ptAntisym} for quasi-primary operators in defining representations.  It merges the bosonic and fermionic cases into a single general object.  Note that in the second equality of \eqref{Eq2ptPS} the hatted projection operator was absorbed by the half-projector of the first quasi-primary operator.  Hence, the proper irreducible representation $\boldsymbol{N}$, with all the traces removed, is obtained through contractions with the half-projectors.  We present two explicit examples below to illustrate this point.


\subsubsection{Example: Symmetric-Traceless Quasi-Primary Operators}

We first turn to the case of quasi-primary operators in symmetric-traceless irreducible representations of the Lorentz group.  The associated hatted projection operator in the embedding space is
\eqna{
(\hat{\mathcal{P}}_{12}^{\ell\boldsymbol{e}_1})_{A_\ell\cdots A_1}^{\phantom{A_\ell\cdots A_1}B_1'\cdots B_\ell'}&=\sum_{i=0}^{\lfloor\ell/2\rfloor}\frac{(-\ell)_{2i}}{2^{2i}i!(-\ell+2-d/2)_i}\A_{12(A_1A_2}\A_{12}^{(B_1'B_2'}\cdots\A_{12A_{2i-1}A_{2i}}\A_{12}^{B_{2i-1}'B_{2i}'}\\
&\phantom{=}\qquad\times\A_{12A_{2i+1}}^{\phantom{12A_{2i+1}}B_{2i+1}'}\cdots\A_{12A_\ell)}^{\phantom{12A_\ell)}B_\ell')},
}
where $\lambda_{\ell\boldsymbol{e}_1}=\sqrt{\ell!/[(d+2\ell-2)(d-1)_{\ell-1}]}$.  Therefore, from \eqref{Eq2ptES} [or directly from \eqref{Eq2ptPS}], we see that the two-point correlation functions are given by
\eqna{
\Vev{\mathcal{O}_i^{(x),\ell\boldsymbol{e}_1}(x_1)\mathcal{O}_j^{(x),\ell\boldsymbol{e}_1}(x_2)}&=\left(\frac{\alpha^2}{2}\right)^\ell(\mathcal{T}^{\ell\boldsymbol{e}_1})^{\mu_1\cdots\mu_\ell}(\mathcal{T}^{\ell\boldsymbol{e}_1})^{\nu_1\cdots\nu_\ell}\\
&\phantom{=}\qquad\times I_{\mu_1\nu_1}(x_1-x_2)\cdots I_{\mu_\ell\nu_\ell}(x_1-x_2)\frac{\lambda_{\ell\boldsymbol{e}_1}\cOPE{}{i}{j}{\1}}{\left[-\frac{1}{2}(x_1-x_2)^2\right]^{\Delta}}.
}[Eq2ptSymTr]
The hatted projection operator implicitly included in the half-projectors,
\eqn{(\hat{\mathcal{P}}^{\ell\boldsymbol{e}_1})_{\mu_\ell\cdots\mu_1}^{\phantom{\mu_\ell\cdots\mu_1}\nu_1'\cdots\nu_\ell'}=\sum_{i=0}^{\lfloor\ell/2\rfloor}\frac{(-\ell)_{2i}}{2^{2i}i!(-\ell+2-d/2)_i}g_{(\mu_1\mu_2}g^{(\nu_1'\nu_2'}\cdots g_{\mu_{2i-1}\mu_{2i}}g^{\nu_{2i-1}'\nu_{2i}'}g_{\mu_{2i+1}}^{\phantom{\mu_{2i+1}}\nu_{2i+1}'}\cdots g_{\mu_\ell)}^{\phantom{\mu_\ell)}\nu_\ell')},}
is the direct equivalent of the embedding space hatted projection operator, which serves to remove the traces in the product of the inversion structure $I_{\mu\nu}(x)$.  Casting \eqref{Eq2ptSymTr} in terms of quasi-primary operators with vector indices clearly leads to the known result for symmetric-traceless quasi-primary operators.


\subsubsection{Example: $\boldsymbol{e}_1+\boldsymbol{e}_r$ Quasi-Primary Operators}

To further demonstrate the use of the formalism in arbitrary representations of the Lorentz group, we consider operators in mixed irreducible representations.  We choose the irreducible representation $\boldsymbol{N}=\boldsymbol{e}_1+\boldsymbol{e}_r$.  The corresponding hatted projector is												   
\eqn{(\hat{\mathcal{P}}_{12}^{\boldsymbol{e}_1+\boldsymbol{e}_r})_{aA}^{\phantom{aA}B'b'}=\A_{12A}^{\phantom{12A}B'}\delta_a^{\phantom{a}b'}-\frac{1}{d}(\Gamma_{12A}\Gamma_{12}^{B'})_a^{\phantom{a}b'},}
in odd spacetime dimensions, with normalization constant $\lambda_{\boldsymbol{e}_1+\boldsymbol{e}_r}=\sqrt{1/[2^{r+1}(d-1)]}$, or
\eqn{(\hat{\mathcal{P}}_{12}^{\boldsymbol{e}_1+\boldsymbol{e}_r})_{\tilde{a}A}^{\phantom{\tilde{a}A}B'\tilde{b}'}=\A_{12A}^{\phantom{12A}B'}\delta_{\tilde{a}}^{\phantom{\tilde{a}}\tilde{b}'}-\frac{1}{d}(\tilde{\Gamma}_{12A}\Gamma_{12}^{B'})_{\tilde{a}}^{\phantom{\tilde{a}}\tilde{b}'},}
in even spacetime dimensions, with normalization constant $\lambda_{\boldsymbol{e}_1+\boldsymbol{e}_r}=\sqrt{1/[2^r(d-1)]}$.

The general result \eqref{Eq2ptPS} then yields
\eqna{
\Vev{\mathcal{O}_i^{(x),\boldsymbol{e}_1+\boldsymbol{e}_r}(x_1)\mathcal{O}_j^{(x),\boldsymbol{e}_1+\boldsymbol{e}_r}(x_2)}&=\frac{\alpha^3}{2}(\mathcal{T}^{\boldsymbol{e}_1+\boldsymbol{e}_r})^{\mu\delta}(\mathcal{T}^{\boldsymbol{e}_1+\boldsymbol{e}_r})^{\nu\epsilon}(x_1-x_2)\cdot(\gamma C^{-1})_{\delta\epsilon}\\
&\phantom{=}\qquad\times I_{\mu\nu}(x_1-x_2)\frac{\lambda_{\boldsymbol{e}_1+\boldsymbol{e}_r}\cOPE{}{i}{j}{\1}}{\left[-\frac{1}{2}(x_1-x_2)^2\right]^{\Delta+1/2}},
}[Eq2pte1erOdd]
in odd spacetime dimensions or
\eqna{
\left.\Vev{\mathcal{O}_i^{(x),\boldsymbol{e}_1+\boldsymbol{e}_r}(x_1)\mathcal{O}_j^{(x),\boldsymbol{e}_1+\boldsymbol{e}_{r-1}}(x_2)}\right|_\text{$r$ even}&=\frac{\alpha^3}{2}(\mathcal{T}^{\boldsymbol{e}_1+\boldsymbol{e}_r})^{\mu\tilde{\delta}}(\mathcal{T}^{\boldsymbol{e}_1+\boldsymbol{e}_{r-1}})^{\nu\epsilon}(x_1-x_2)\cdot(\tilde{\gamma}C^{-1})_{\tilde{\delta}\epsilon}\\
&\phantom{=}\qquad\times I_{\mu\nu}(x_1-x_2)\frac{\lambda_{\boldsymbol{e}_1+\boldsymbol{e}_r}\cOPE{}{i}{j}{\1}}{\left[-\frac{1}{2}(x_1-x_2)^2\right]^{\Delta+1/2}},\\
\left.\Vev{\mathcal{O}_i^{(x),\boldsymbol{e}_1+\boldsymbol{e}_r}(x_1)\mathcal{O}_j^{(x),\boldsymbol{e}_1+\boldsymbol{e}_r}(x_2)}\right|_\text{$r$ odd}&=\frac{\alpha^3}{2}(\mathcal{T}^{\boldsymbol{e}_1+\boldsymbol{e}_r})^{\mu\tilde{\delta}}(\mathcal{T}^{\boldsymbol{e}_1+\boldsymbol{e}_r})^{\nu\tilde{\epsilon}}(x_1-x_2)\cdot(\tilde{\gamma}C^{-1})_{\tilde{\delta}\tilde{\epsilon}}\\
&\phantom{=}\qquad\times I_{\mu\nu}(x_1-x_2)\frac{\lambda_{\boldsymbol{e}_1+\boldsymbol{e}_r}\cOPE{}{i}{j}{\1}}{\left[-\frac{1}{2}(x_1-x_2)^2\right]^{\Delta+1/2}},
}[Eq2pte1erEven]
in even spacetime dimensions.  As expected, \eqref{Eq2pte1erOdd} and \eqref{Eq2pte1erEven} are simply built from the results of the appropriate defining representations and are then properly constrained to the right irreducible representation by removing traces using
\eqn{(\hat{\mathcal{P}}^{\boldsymbol{e}_1+\boldsymbol{e}_r})_{\delta\mu}^{\phantom{\delta\mu}\mu'\delta'}=g_\mu^{\phantom{\mu}\mu'}\delta_\delta^{\phantom{\delta}\delta'}-\frac{1}{d}(\gamma_\mu\gamma^{\mu'})_\delta^{\phantom{\delta}\delta'},}
in odd spacetime dimensions or 
\eqn{(\hat{\mathcal{P}}^{\boldsymbol{e}_1+\boldsymbol{e}_r})_{\tilde{\delta}\mu}^{\phantom{\tilde{\delta}\mu}\mu'\tilde{\delta}'}=g_\mu^{\phantom{\mu}\mu'}\delta_{\tilde{\delta}}^{\phantom{\tilde{\delta}}\tilde{\delta}'}-\frac{1}{d}(\gamma_\mu\tilde{\gamma}^{\mu'})_{\tilde{\delta}}^{\phantom{\tilde{\delta}}\tilde{\delta}'},}
in even spacetime dimensions.  We remark that the embedding space normalization constants are enhanced with respect to the corresponding ones in position space by a factor of two, as explained earlier.


\subsection{Conformal Covariance}

To verify that the two-point correlation functions \eqref{Eq2ptPS} are indeed correct, it is sufficient to check their covariance under conformal transformations.  Ascertaining covariance under both translation and dilatation is effortless.  Covariance under Lorentz transformations is also easy to verify since
\eqn{(\sigma_{\mu\nu}\mathcal{T}^{\boldsymbol{N}})=(\mathcal{T}^{\boldsymbol{N}}s_{\mu\nu}),\qquad(\mathcal{L}_1+\mathcal{L}_2)_{\mu\nu}(x_1-x_2)^\lambda=-[s_{\mu\nu}(x_1-x_2)]^\lambda.}
Here the index-free notation of \cite{Fortin:2019fvx,Fortin:2019dnq} has been used.  The only non-trivial transformations left to verify are the special conformal transformations.

We first apply translational invariance to shift one of the two spacetime points to the origin, which allows us to recast the two-point correlation functions \eqref{Eq2ptPS} as 
\eqn{\Vev{\mathcal{O}_i^{(x)\boldsymbol{N}}(x)\mathcal{O}_j^{(x)\boldsymbol{N}^C}(0)}=\left(\frac{\alpha^2}{2}\right)^{S-\xi}(\mathcal{T}^{\boldsymbol{N}})^{\{\mu\delta\}}(\mathcal{T}^{\boldsymbol{N}^C})^{\{\nu\epsilon\}}[\alpha x\cdot(\gamma C^{-1})_{\delta\epsilon}]^{2\xi}[I_{\mu\nu}(x)]^{n_v}\frac{\lambda_{\boldsymbol{N}}\cOPE{}{i}{j}{\1}}{\left(-\frac{1}{2}x^2\right)^{\Delta+\xi}}.}
Since the special conformal generators annihilate quasi-primary operators at the origin of spacetime, covariance under special conformal transformations is equivalent to
\eqn{-i\left(2x_\mu x\cdot\partial-x^2\partial_\mu+2\Delta x_\mu\right)\Vev{\mathcal{O}_i^{(x)\boldsymbol{N}}(x)\mathcal{O}_j^{(x)\boldsymbol{N}^C}(0)}=2x^\nu\Vev{(\sigma_{\mu\nu}\mathcal{O}_i^{(x)\boldsymbol{N}})(x)\mathcal{O}_j^{(x)\boldsymbol{N}^C}(0)},}
which is ensured by noting that
\eqn{
\begin{gathered}
-i(2x_\mu x\cdot\partial-x^2\partial_\mu)(\alpha x\cdot\gamma C^{-1})^{2\xi}=-2i\xi x_\mu(\alpha x\cdot\gamma C^{-1})^{2\xi}+2\xi x^\rho[\alpha(s_{\mu\rho}x)\cdot\gamma C^{-1}]^{2\xi},\\
-i(2x_\mu x\cdot\partial-x^2\partial_\mu)I_{\nu\lambda}(x)=x^\rho(s_{\mu\rho}I)_{\nu\lambda}(x),\\
-i(2x_\mu x\cdot\partial-x^2\partial_\mu)\frac{1}{\left(-\frac{1}{2}x^2\right)^{\Delta+\xi}}=2i(\Delta+\xi)x_\mu\frac{1}{\left(-\frac{1}{2}x^2\right)^{\Delta+\xi}}.
\end{gathered}
}
Hence, we find that the two-point correlation functions \eqref{Eq2ptPS} are indeed covariant under conformal transformations, furnishing a first sanity check on the consistency of the embedding space formalism.


\section{Unitarity Conditions}\label{SecUnitarity}

In a unitary CFT, two-point correlation functions must satisfy the Wightman positivity condition \cite{Streater:1989vi,Fradkin:1996is}.  Usually, unitarity is verified from the correlation functions in Euclidean signature through reflection-positivity, using radial quantization (see \textit{e.g.} \cite{Rychkov:2016iqz}).  Since all computations in this work are performed in Lorentzian signature, the Wightman positivity condition can be applied directly to obtain the unitarity conditions.


\subsection{A Metric in the Space of Quasi-Primary Operators}

As expressed in \eqref{Eq2ptESDummy}, two-point correlation functions are non-vanishing exclusively between quasi-primary operators in conjugate representations with respect to each other.  It is therefore convenient to use this property directly to rewrite \eqref{Eq2ptESDummy} as follows:
\eqn{\Vev{\mathcal{O}_i(\eta_1)\mathcal{O}_j^C(\eta_2)}=(-1)^{\xi r(r+3)}(\mathcal{T}_{12}^{\boldsymbol{N}}\Gamma)^{\{Aa\}}(\mathcal{T}_{21}^{\boldsymbol{N}^C}\Gamma)^{\{Bb\}}[-i(C_\Gamma^{-1})_{ab}]^{2\xi}(-g_{AB})^{n_v}\frac{\lambda_{\boldsymbol{N}}c_{ij}}{\ee{1}{2}{\tau}},}[Eq2ptESUnitarity]
where $\boldsymbol{N}_i=\boldsymbol{N}_j=\boldsymbol{N}$ and $c_{ij}$ is a new OPE coefficient matrix, which will be constrained and reinterpreted shortly.  For future convenience, we introduced a phase in \eqref{Eq2ptESUnitarity} that differs from the choice made in Section \ref{SecGeneral}.

One constraint on the matrix $c_{ij}$ can be derived by considering the complex conjugated two-point correlation functions and demanding that they match the original form.\footnote{Consistency condition implies that the complex conjugate of a product of Grassmann variables $(\alpha\beta)^*$ corresponds to the product of the complex conjugate Grassmann variables in inverted order $\beta^*\alpha^*$.}  Specifically,
\eqna{
\Vev{\mathcal{O}_i(\eta_1)\mathcal{O}_j^C(\eta_2)}&=\Vev{\mathcal{O}_j^{C*}(\eta_2)\mathcal{O}_i^*(\eta_1)}^*=\Vev{(B_\Gamma^{-*}\mathcal{O}_j)(\eta_2)(B_\Gamma\mathcal{O}_i^C)(\eta_1)}^*\\
&=\left[(-1)^{\xi r(r+3)}(B_\Gamma^{-*}\mathcal{T}_{21}^{\boldsymbol{N}}\Gamma)\cdot(B_\Gamma\mathcal{T}_{12}^{\boldsymbol{N}^C}\Gamma)\frac{\lambda_{\boldsymbol{N}}c_{ji}}{\ee{1}{2}{\tau}}\right]^*\\
&=(-1)^{\xi r(r+3)}(\mathcal{T}_{21}^{\boldsymbol{N}^C}\Gamma)^{\{Aa\}}(\mathcal{T}_{12}^{\boldsymbol{N}}\Gamma)^{\{Bb\}}[i(B_\Gamma^{-1}C_\Gamma^{-*}B_\Gamma^{-1})_{ab}]^{2\xi}(-g_{AB})^{n_v}\frac{\lambda_{\boldsymbol{N}}c_{ji}^*}{\ee{1}{2}{\tau*}},
}[Eq2ptESUnitarityS]
where the Lorentzian signature property $B_\Gamma^{-1}\hat{\mathcal{P}}_{21}^{\boldsymbol{N}*}=\hat{\mathcal{P}}_{21}^{\boldsymbol{N}^C}B_\Gamma^{-1}$ was used in the last identity.  Because $B_\Gamma^{-1}C_\Gamma^{-*}B_\Gamma^{-1}=-C_\Gamma^{-T}$ in Lorentzian signature, we find that the constraint obtained by comparing \eqref{Eq2ptESUnitarity} and \eqref{Eq2ptESUnitarityS} simply corresponds to
\eqn{c_{ji}^*=c_{ij}.}[EqMetricS]
Hence, we find that for all quasi-primary operators, the matrix $c_{ij}$ is Hermitian with real eigenvalues $c_i$.  Note that $\ee{1}{2}{\tau}$ was assumed to be real.  This last convention, which will be discussed in more detail later, is used to simplify the unitarity conditions.  Indeed, for all quasi-primary operators, the unitarity conditions constrain the sign of the real eigenvalues $c_i$ of the matrix $c_{ij}$, making the latter a metric in the space of all quasi-primary operators.

Another constraint on the matrix $c_{ij}$ can be derived by considering the two-point function of self-conjugate quasi-primary operators $\mathcal{O}_i^C(\eta)=\mathcal{O}_i(\eta)$.  Applying the OPE to the product of permuted quasi-primary operators results in a different expression for two-point correlation functions given by
\begingroup\makeatletter\def\f@size{11}\check@mathfonts\def\maketag@@@#1{\hbox{\m@th\large\normalfont#1}}%
\eqna{
\Vev{\mathcal{O}_i(\eta_1)\mathcal{O}_j(\eta_2)}&=(-1)^{2\xi}\Vev{\mathcal{O}_j(\eta_2)\mathcal{O}_i(\eta_1)}=(-1)^{\xi(r+1)(r+2)}(\mathcal{T}_{21}^{\boldsymbol{N}^C}\Gamma)(\mathcal{T}_{12}^{\boldsymbol{N}}\Gamma)\cdot\frac{\lambda_{\boldsymbol{N}}c_{ji}\hat{\mathcal{P}}_{21}^{\boldsymbol{N}^C}}{\ee{1}{2}{\tau}}\\
&=(\mathcal{T}_{21}^{\boldsymbol{N}^C}\Gamma)^{\{Aa\}}(\mathcal{T}_{12}^{\boldsymbol{N}}\Gamma)^{\{Bb\}}(\hat{\mathcal{P}}_{21}^{\boldsymbol{N}^C})_{\{aA\}}^{\phantom{\{aA\}}\{B'b'\}}\\
&\phantom{=}\qquad\times[-i(C_\Gamma^{-1})_{b'b}]^{2\xi} (-g_{B'B})^{n_v}\frac{(-1)^{\xi(r+1)(r+2)}\lambda_{\boldsymbol{N}}c_{ji}}{\ee{1}{2}{\tau}}.
}[Eq2ptESUnitarityT]
\endgroup
Intuitively, we expect the two distinct expressions \eqref{Eq2ptESUnitarity} and \eqref{Eq2ptESUnitarityT} for the two-point functions to obviously match.  This observation yields another constraint on the matrix $c_{ij}$, which can be obtained via the application of the identity,\footnote{The identity \eqref{EqPC} originates from the OPE and states that two-point correlation functions are non-vanishing for quasi-primary operators in contragredient-reflected representations of one another \cite{Fortin:2019fvx,Fortin:2019dnq}.}
\eqn{(\hat{\mathcal{P}}_{12}^{\boldsymbol{N}})_{\{aA\}}^{\phantom{\{aA\}}\{B'b'\}}[(C_\Gamma^{-1})_{b'b}]^{2\xi}(g_{B'B})^{n_v}=[(C_\Gamma^{-1})_{ab'}]^{2\xi}(g_{AB'})^{n_v}(\hat{\mathcal{P}}_{21}^{\boldsymbol{N}^C})_{\{bB\}}^{\phantom{\{bB\}}\{B'b'\}},}[EqPC]
and $C_\Gamma^T=(-1)^{(r+1)(r+2)/2}C_\Gamma$.  The resulting constraint is given by
\eqn{c_{ji}=(-1)^{\xi r(r+3)}c_{ij}=c_{ij}.}[EqMetricT]
The fact that fermionic quasi-primary operators that are self-conjugate (\textit{i.e.} when the Majorana condition can be imposed) exist only for $r=0,1$ was used in the last identity.  There is no analog of the above constraint for quasi-primary operators that are not self-conjugate.

The equivalent constraints can be obtained directly in position space.  First, we observe that in \eqref{Eq2ptPS}, all quantities are position-space quantities, with the exception of $\alpha=\pm1,\pm i$, which was introduced to show that the Majorana condition can be imposed in embedding space if and only if it can be imposed in position space (see Section 3 in \cite{Fortin:2019dnq}).  We are free to now fix $\alpha=1$ without loss of generality, irrespective of the Majorana condition.  Hence, the two-point correlation functions \eqref{Eq2ptESUnitarity} in position space are given by
\eqna{
\Vev{\mathcal{O}_i^{(x)\boldsymbol{N}}(x_1)\mathcal{O}_j^{(x)\boldsymbol{N}C}(x_2)}&=(-1)^{\xi r(r+3)}\left(\frac{1}{2}\right)^{S-\xi}(\mathcal{T}^{\boldsymbol{N}})^{\{\mu\delta\}}(\mathcal{T}^{\boldsymbol{N}^C})^{\{\nu\epsilon\}}\\
&\phantom{=}\quad\times[-i(x_1-x_2)\cdot(\gamma C^{-1})_{\delta\epsilon}]^{2\xi}[-I_{\mu\nu}(x_1-x_2)]^{n_v}\frac{\lambda_{\boldsymbol{N}}c_{ij}}{\left[-\frac{1}{2}(x_1-x_2)^2\right]^{\Delta+\xi}},
}[Eq2ptPSUnitarity]
with the understanding that $\alpha=1$.

It is now a trivial matter to use \eqref{Eq2ptPSUnitarity} to demonstrate that the matrices $c_{ij}$ satisfy \eqref{EqMetricS} and \eqref{EqMetricT} directly in position space, assuming a space-like interval.  This observation explains the choice $\ee{1}{2}{\tau}\in\mathbb{R}$ made previously.\footnote{In fact, from \eqref{EqetaProd}, the proper choice is $(-\eta_1^{d+1}+\eta_1^{d+2})(-\eta_2^{d+1}+\eta_2^{d+2})\in\mathbb{R}^+$ to avoid superfluous phases.  In any case, this prefactor is absorbed when quasi-primary operators are projected back to position space.}


\subsection{Positivity}

For the purpose of analyzing the Wightman positivity condition, we are specifically interested in the two-point correlation functions
\eqna{
\Vev{\mathcal{O}_i^{(x)\boldsymbol{N}}(x_1)\mathcal{O}_j^{(x)\boldsymbol{N}*}(x_2)}&=(-1)^{f(\boldsymbol{N})}\Vev{\mathcal{O}_i^{(x)\boldsymbol{N}}(x_1)B\mathcal{O}_j^{(x)\boldsymbol{N}C}(x_2)}\\
&=(-1)^{\xi r(r+3)}\left(\frac{1}{2}\right)^{S-\xi}(\mathcal{T}^{\boldsymbol{N}})^{\{\mu\delta\}}(\mathcal{T}^{\boldsymbol{N}*})_{\{\nu\epsilon\}}\\
&\phantom{=}\quad\times[-i(x_1-x_2)\cdot(\gamma C^{-1}B^T)_\delta^{\phantom{\delta}\epsilon}]^{2\xi}[-I_\mu^{\phantom{\mu}\nu}(x_1-x_2)]^{n_v}\frac{\lambda_{\boldsymbol{N}}c_{ij}}{\left[-\frac{1}{2}(x_1-x_2)^2\right]^{\Delta+\xi}}\\
&=\left(\frac{1}{2}\right)^{S-\xi}(\mathcal{T}^{\boldsymbol{N}})^{\{\mu\delta\}}(\mathcal{T}^{\boldsymbol{N}*})_{\{\nu\epsilon\}}\\
&\phantom{=}\quad\times[-i(x_1-x_2)\cdot(\gamma A)_\delta^{\phantom{\delta}\epsilon}]^{2\xi}[-I_\mu^{\phantom{\mu}\nu}(x_1-x_2)]^{n_v}\frac{\lambda_{\boldsymbol{N}}c_{ij}}{\left[-\frac{1}{2}(x_1-x_2)^2\right]^{\Delta+\xi}},
}
where
\eqn{f(\boldsymbol{N})=(r+1)\times\begin{cases}\sum_{i=1}^{r-1}(i+1)N_i+(r+1)\lfloor N_r/2\rfloor&d\text{ odd}\\\sum_{i=1}^{r-2}(i+1)N_i+r\,\text{min}\{N_{r-1},N_r\}+(r+1)\lfloor|N_{r-1}-N_r|/2\rfloor&d\text{ even}\end{cases},}
and $A=\gamma^0$ in Lorentzian signature.  Here the definition $\mathcal{O}_i^{(x)\boldsymbol{N}C}(x)=(-1)^{f(\boldsymbol{N})}B^{-1}\mathcal{O}_i^{(x)\boldsymbol{N}*}(x)$ for the conjugate quasi-primary operators was made such that self-conjugate quasi-primary operators satisfy $\mathcal{O}_i^{(x)\boldsymbol{N}C}(x)=\mathcal{O}_i^{(x)\boldsymbol{N}}(x)$.

We now smear the quasi-primary operators with a suitable finite set of test functions $h_i(x)$ (which are infinitely differentiable and vanish outside some bounded region of spacetime) as in
\eqn{\mathcal{O}_i^{(x)\boldsymbol{N}}(x)\to\mathcal{O}^{(x)}(h)=\sum_i\int d^dx\,h_i(x)*\mathcal{O}_i^{(x)\boldsymbol{N}}(x),}
where the star product corresponds to the complete contraction of the spinor indices.  This brings us to the Wightman positivity condition
\eqn{\sum_{i,j}\int d^dx_1d^dx_2\,\Vev{[h_i*\mathcal{O}_i^{(x)\boldsymbol{N}}](x_1)[h_j*\mathcal{O}_j^{(x)\boldsymbol{N}}]^*(x_2)}\geq0,}
for all suitable test functions.  Upon choosing $h_i(x)=g_i(x)\cdot\mathcal{T}_{\boldsymbol{N}}$, where again the dot product corresponds to the full contraction of the dummy indices, the Wightman positivity condition assumes the form
\eqna{
&\sum_{i,j}\int d^dx_1d^dx_2\,\left(\frac{1}{2}\right)^{S-\xi}(g_i(x_1)\cdot\hat{\mathcal{P}}^{\boldsymbol{N}})^{\{\mu\delta\}}(g_j^*(x_2)\cdot\hat{\mathcal{P}}^{\boldsymbol{N}*})_{\{\nu\epsilon\}}\\
&\qquad\times[-i(x_1-x_2)\cdot(\gamma\gamma^0)_\delta^{\phantom{\delta}\epsilon}]^{2\xi}[- I_\mu^{\phantom{\mu}\nu}(x_1-x_2)]^{n_v}\frac{\lambda_{\boldsymbol{N}}c_{ij}}{\left[-\frac{1}{2}(x_1-x_2)^2\right]^{\Delta+\xi}}\geq0.
}[EqWP]
This last result \eqref{EqWP}, which is commonly written as
\eqn{\sum_{i,j}\int d^dx_1d^dx_2\,g_i(x_1)\cdot W_{ij}(x_1,x_2)\cdot g_j^*(x_2)\geq0,}[EqWPx]
is the usual Wightman positivity condition for two-point correlation functions.  The condition \eqref{EqWPx} demands that the $W_{ij}(x_1,x_2)$ be Wightman functions, which is easily implemented by introducing the usual $i\epsilon$-prescription $x^2\to x^2-i\epsilon x^0$ where $x^2$ is the norm of $x^\mu$ while $x^0$ is the time component of $x^\mu$, \textit{i.e.} $-\frac{1}{2}(x_1-x_2)^2\to-\frac{1}{2}[(x_1-x_2)^2-i\epsilon(x_1^0-x_2^0)]$.  The $i\epsilon$-prescription must also be imposed on $I_\mu^{\phantom{\mu}\nu}(x_1-x_2)$.

Upon taking the Fourier transform of \eqref{EqWPx} using the fact that $W_{ij}(x_1,x_2)=W_{ij}(x_1-x_2)$, we are led to the Wightman positivity condition
\eqn{\sum_{i,j}\int\frac{d^dp}{(2\pi)^d}\,\widetilde{g_i}(-p)\cdot\widetilde{W}_{ij}(p)\cdot\widetilde{g_j^*}(p)\geq0,}[EqWPp]
in momentum space.  Here $\widetilde{g_j^*}(p)$ is the Fourier transform of $g_j^*(x)$, while $\tilde{g}_j^*(p)$ is the complex conjugate of the Fourier transform of $g_j(x)$.  Hence, for a set of real test functions $g_i(x)$ one has $\widetilde{g_i^*}(-p)=\tilde{g}_i^*(p)$.  Moreover, $\widetilde{g_i}(p)$ decreases sufficiently quickly as $p^2\to\infty$.

Since the Fourier transform of the scalar Wightman function is given by
\eqn{\widetilde{W}_\Delta(p)=\int d^dx\,\frac{e^{ip\cdot x}}{\left[-\frac{1}{2}(x^2-i\epsilon x^0)\right]^\Delta}=\frac{2^{d+1-\Delta}\pi^{d/2+1}}{\Gamma(\Delta)\Gamma(\Delta+1-d/2)}\theta(p^0)\theta(p^2)(p^2)^{\Delta-d/2},}[EqFW]
we see that the Wightman positivity condition in momentum space \eqref{EqWPp} with the general Wightman function \eqref{EqWPx} and \eqref{EqFW} translates to the statement
\eqna{
&\sum_{i,j}\int\frac{d^dp}{(2\pi)^d}\,\left(\frac{1}{2}\right)^{S-\xi}\lambda_{\boldsymbol{N}}c_{ij}(\widetilde{g_i}(-p)\cdot\hat{\mathcal{P}}^{\boldsymbol{N}})^{\{\mu\delta\}}(\tilde{g}_j^*(-p)\cdot\hat{\mathcal{P}}^{\boldsymbol{N}*})_{\{\nu\epsilon\}}\\
&\qquad\times[\partial\cdot(\gamma\gamma^0)_\delta^{ \phantom{\delta}\epsilon}]^{2\xi}\left[\partial_\mu\partial^\nu-\delta_\mu^{\phantom{\mu}\nu}\frac{\partial^2}{2}\right]^{n_v}\widetilde{W}_{\Delta+n_v+\xi}(p)\geq0,
}
for all real test functions $g_i(x)$.  By choosing smearing functions centered around $p^0=E>0$ and $\boldsymbol{p}=\boldsymbol{0}$ with widths $\Delta E$ and $\Delta p$, respectively, as in
\eqn{\widetilde{g_i}(p)=e^{-\frac{(p^0-E)^2}{2\Delta E}-\frac{|\boldsymbol{p}|^2}{2\Delta p}}\zeta_i,}
where $\zeta_i$ are polarization tensors, the last result translates to the statement
\eqna{
&\sum_{i,j}c_{ij}(\zeta_i\cdot\hat{\mathcal{P}}^{\boldsymbol{N}})^{\{\mu\delta\}}(\zeta_j^*\cdot\hat{\mathcal{P}}^{\boldsymbol{N}*})_{\{\nu\epsilon\}}\int\frac{d^dp}{(2\pi)^d}\,e^{-\frac{(p^0-E)^2}{\Delta E}-\frac{|\boldsymbol{p}|^2}{\Delta p}}\\
&\qquad\times[\partial\cdot(\gamma\gamma^0)_\delta^{ \phantom{\delta}\epsilon}]^{2\xi}\left[\partial_\mu\partial^\nu-\delta_\mu^{\phantom{\mu}\nu}\frac{\partial^2}{2}\right]^{n_v}\widetilde{W}_{\Delta+n_v+\xi}(p)\geq0.
}
For very small widths $\Delta E\to0$ and $\Delta p\to0$, this becomes
\eqn{\sum_{i,j}\frac{c_{ij}(\zeta_i\cdot\hat{\mathcal{P}}^{\boldsymbol{N}})^{\{\mu\delta\}}(\zeta_j^*\cdot\hat{\mathcal{P}}^{\boldsymbol{N}*})_{\{\nu\epsilon\}}}{\Gamma(\Delta+n_v+\xi)\Gamma(\Delta+n_v+\xi+1-d/2)}\left.[\partial\cdot(\gamma\gamma^0)_\delta^{\phantom{\delta}\epsilon}]^{2\xi}\left[\partial_\mu\partial^\nu-\delta_\mu^{\phantom{\mu}\nu}\frac{\partial^2}{2}\right]^{n_v}(p^2)^{\Delta+n_v+\xi}\right|_{\substack{p^0\to E\\\boldsymbol{p}\to\boldsymbol{0}}}\geq0,}[EqWPfinal]
after using \eqref{EqFW}.  In \eqref{EqWPfinal}, it is understood that the momentum derivatives are performed first, followed by the substitutions $p^0\to E$ and $\boldsymbol{p}=\boldsymbol{0}$.

By appropriately choosing the polarization tensors in \eqref{EqWPfinal}, we find different constraints on the matrix $c_{ij}$ and the conformal dimension $\Delta$.  Together, they imply that in a unitary CFT, the matrix $c_{ij}$ is positive semi-definite with real non-negative eigenvalues $c_i$,
\eqn{c_{ij}\to c_i\delta_{ij}\text{ with }c_i\geq0,}[EqUnitarity]
and can thus be understood as a metric in the space of quasi-primary operators, as stated above.\footnote{For a vanishing eigenvalue $c_i=0$, the corresponding quasi-primary operator $\mathcal{O}_i^{(x)\boldsymbol{N}}$ decouples from the theory.}  Moreover, the conformal dimensions satisfy the proper unitarity bounds on the associated irreducible representation.  This last observation is usually obtained by considering descendants.  Here, it can be seen from the smearing around a sharp region in momentum space, which corresponds to a broad spacetime region.  Hence \eqref{EqWPfinal} includes descendants.

For example, for vector quasi-primary operators, taking all polarization vectors to vanish except the $i$-th one, \eqref{EqWPfinal} leads to
\eqn{c_{ii}\frac{(\Delta+1-d)|\zeta^0|^2+(\Delta-1)|\boldsymbol{\zeta}|^2}{\Gamma(\Delta+1)\Gamma(\Delta+1-d/2)}\geq0,}
which implies that $c_{ii}$ is non-negative and $\Delta\geq d-1$, as expected from unitarity.


\section{Conclusion}\label{SecConc}

We have explicitly computed the most general two-point function of quasi-primary operators in arbitrary Lorentz representations using the recent embedding space formalism \cite{Fortin:2019fvx,Fortin:2019dnq}.  The complete result is specified in \eqref{Eq2ptESDummy} and its corresponding projection to position space is given in \eqref{Eq2ptPS}.  We have performed several checks of the formalism by explicitly taking the embedding space results and projecting them to position space.  In all cases, we have found that the form of the results matches expectations from conformal covariance.  Moreover, we have directly verified that the most general expression for the two-point function is covariant under the full conformal group, thus confirming its validity.

In addition, we have studied constraints on the OPE coefficient matrix $c_{ij}$ that arise from considering the complex conjugate of the correlator.  Furthermore, because the embedding space OPE is inherently not symmetric in the operator ordering, we have examined the implications of interchanging the operator order on the coefficients.  Obviously, the lack of symmetry is spurious, implying symmetries of the OPE coefficient matrices appearing in the two-point functions.  The respective results are summarized in \eqref{EqMetricS} and \eqref{EqMetricT}.  Lastly, we have explored unitarity conditions on generic quasi-primary operators.  These constrain the signs of the eigenvalues of the OPE matrices \eqref{EqUnitarity}.

It is clearly of interest to determine the general form of the two-point functions, as it contains some essential physical ingredients necessary for the understanding of the higher-point functions in the newly developed formalism.  The projection operators, which encode all the essential group theoretic information, appear on the same footing in the construction of three-point, four-point, and general $M$-point functions.  Further, the two-point functions encode the simplest unitarity constraints in a given theory.
 
This work is a first step in the application of the program of computing the most general $M$-point correlation functions in the context of this formalism.  In upcoming publications, we will proceed to construct general expressions for three-point functions of quasi-primary operators in generic Lorentz representations and then provide results for four-point functions.  Much exciting work lies ahead, and we anticipate that exploiting this formalism further will eventually shed considerable light on the space of conformal field theories.


\ack{
The authors would like to thank Simon Caron-Huot and Wenjie Ma for useful discussions.  The work of JFF and VP is supported by NSERC and FRQNT.
}


\bibliography{TwoPtFcts}

\begin{thebibliography}{10}
\ifx\href\asklfhas\newcommand{\href}[2]{#2}\fi
\ifx\arxivref\asklfhas\newcommand{\arxivref}[2]{\href{http://arxiv.org/abs/#1}{#2}}\fi
\ifx\doiref\asklfhas\newcommand{\doiref}[2]{\href{http://dx.doi.org/#1}{#2}}\fi
\parskip 0pt
\normalsize

\bibitem{Fortin:2019fvx}
J.-F. Fortin \& W.~Skiba,
\textit{``{A recipe for conformal blocks}''},
\normalsize{\texttt{\arxivref{1905.00036}{arXiv:1905.00036}}}\ignorespaces
\bibitem{Fortin:2019dnq}
J.-F. Fortin \& W.~Skiba,
\textit{``{New Methods for Conformal Correlation Functions}''},
\normalsize{\texttt{\arxivref{1905.00434}{arXiv:1905.00434}}}\ignorespaces
\bibitem{Dolan:2000ut}
F.~A. Dolan \& H.~Osborn,
\textit{``{Conformal four point functions and the operator product
  expansion}''},
\doiref{10.1016/S0550-3213(01)00013-X}{Nucl.~Phys. \textbf{B599}, 459
  (2001)\ignorespaces}\ignorespaces,
\normalsize{\texttt{\arxivref{hep-th/0011040}{hep-th/0011040}}}\ignorespaces
\bibitem{Dolan:2003hv}
F.~A. Dolan \& H.~Osborn,
\textit{``{Conformal partial waves and the operator product expansion}''},
\doiref{10.1016/j.nuclphysb.2003.11.016}{Nucl.~Phys. \textbf{B678}, 491
  (2004)\ignorespaces}\ignorespaces,
\normalsize{\texttt{\arxivref{hep-th/0309180}{hep-th/0309180}}}\ignorespaces
\bibitem{Dolan:2011dv}
F.~A. Dolan \& H.~Osborn,
\textit{``{Conformal Partial Waves: Further Mathematical Results}''},
\normalsize{\texttt{\arxivref{1108.6194}{arXiv:1108.6194}}}\ignorespaces
\bibitem{Rattazzi:2008pe}
R.~Rattazzi, V.~S. Rychkov, E.~Tonni \& A.~Vichi,
\textit{``{Bounding scalar operator dimensions in 4D CFT}''},
\doiref{10.1088/1126-6708/2008/12/031}{JHEP \textbf{0812}, 031
  (2008)\ignorespaces}\ignorespaces,
\normalsize{\texttt{\arxivref{0807.0004}{arXiv:0807.0004}}}\ignorespaces
\bibitem{Ferrara:1973yt}
S.~Ferrara, A.~F. Grillo \& R.~Gatto,
\textit{``{Tensor representations of conformal algebra and conformally
  covariant operator product expansion}''},
\doiref{10.1016/0003-4916(73)90446-6}{Annals~Phys. \textbf{76}, 161
  (1973)\ignorespaces}\ignorespaces
\bibitem{Polyakov:1974gs}
A.~M. Polyakov,
\textit{``{Nonhamiltonian approach to conformal quantum field theory}''},
Zh.~Eksp.~Teor.~Fiz. \textbf{66}, 23 (1974)\ignorespaces\ignorespaces,
[Sov. Phys. JETP39,9(1974)]\ignorespaces
\bibitem{Poland:2018epd}
D.~Poland, S.~Rychkov \& A.~Vichi,
\textit{``{The Conformal Bootstrap: Theory, Numerical Techniques, and
  Applications}''},
\doiref{10.1103/RevModPhys.91.015002}{Rev.~Mod.~Phys. \textbf{91}, 15002
  (2019)\ignorespaces}\ignorespaces,
\normalsize{\texttt{\arxivref{1805.04405}{arXiv:1805.04405}}}\ignorespaces,
[Rev. Mod. Phys.91,015002(2019)]\ignorespaces
\bibitem{Dirac:1936fq}
P.~A.~M. Dirac,
\textit{``{Wave equations in conformal space}''},
\doiref{10.2307/1968455}{Annals~Math. \textbf{37}, 429
  (1936)\ignorespaces}\ignorespaces
\bibitem{Mack:1976pa}
G.~Mack,
\textit{``{Convergence of Operator Product Expansions on the Vacuum in
  Conformal Invariant Quantum Field Theory}''},
\doiref{10.1007/BF01609130}{Commun.~Math.~Phys. \textbf{53}, 155
  (1977)\ignorespaces}\ignorespaces
\bibitem{Rattazzi:2010gj}
R.~Rattazzi, S.~Rychkov \& A.~Vichi,
\textit{``{Central Charge Bounds in 4D Conformal Field Theory}''},
\doiref{10.1103/PhysRevD.83.046011}{Phys.~Rev. \textbf{D83}, 046011
  (2011)\ignorespaces}\ignorespaces,
\normalsize{\texttt{\arxivref{1009.2725}{arXiv:1009.2725}}}\ignorespaces
\bibitem{Poland:2010wg}
D.~Poland \& D.~Simmons-Duffin,
\textit{``{Bounds on 4D Conformal and Superconformal Field Theories}''},
\doiref{10.1007/JHEP05(2011)017}{JHEP \textbf{1105}, 017
  (2011)\ignorespaces}\ignorespaces,
\normalsize{\texttt{\arxivref{1009.2087}{arXiv:1009.2087}}}\ignorespaces
\bibitem{Costa:2011mg}
M.~S. Costa, J.~Penedones, D.~Poland \& S.~Rychkov,
\textit{``{Spinning Conformal Correlators}''},
\doiref{10.1007/JHEP11(2011)071}{JHEP \textbf{1111}, 071
  (2011)\ignorespaces}\ignorespaces,
\normalsize{\texttt{\arxivref{1107.3554}{arXiv:1107.3554}}}\ignorespaces
\bibitem{SimmonsDuffin:2012uy}
D.~Simmons-Duffin,
\textit{``{Projectors, Shadows, and Conformal Blocks}''},
\doiref{10.1007/JHEP04(2014)146}{JHEP \textbf{1404}, 146
  (2014)\ignorespaces}\ignorespaces,
\normalsize{\texttt{\arxivref{1204.3894}{arXiv:1204.3894}}}\ignorespaces
\bibitem{Hartman:2016dxc}
T.~Hartman, S.~Jain \& S.~Kundu,
\textit{``{A New Spin on Causality Constraints}''},
\doiref{10.1007/JHEP10(2016)141}{JHEP \textbf{1610}, 141
  (2016)\ignorespaces}\ignorespaces,
\normalsize{\texttt{\arxivref{1601.07904}{arXiv:1601.07904}}}\ignorespaces
\bibitem{Hofman:2016awc}
D.~M. Hofman, D.~Li, D.~Meltzer, D.~Poland \& F.~Rejon-Barrera,
\textit{``{A Proof of the Conformal Collider Bounds}''},
\doiref{10.1007/JHEP06(2016)111}{JHEP \textbf{1606}, 111
  (2016)\ignorespaces}\ignorespaces,
\normalsize{\texttt{\arxivref{1603.03771}{arXiv:1603.03771}}}\ignorespaces
\bibitem{Hartman:2016lgu}
T.~Hartman, S.~Kundu \& A.~Tajdini,
\textit{``{Averaged Null Energy Condition from Causality}''},
\doiref{10.1007/JHEP07(2017)066}{JHEP \textbf{1707}, 066
  (2017)\ignorespaces}\ignorespaces,
\normalsize{\texttt{\arxivref{1610.05308}{arXiv:1610.05308}}}\ignorespaces
\bibitem{Afkhami-Jeddi:2016ntf}
N.~Afkhami-Jeddi, T.~Hartman, S.~Kundu \& A.~Tajdini,
\textit{``{Einstein gravity 3-point functions from conformal field theory}''},
\doiref{10.1007/JHEP12(2017)049}{JHEP \textbf{1712}, 049
  (2017)\ignorespaces}\ignorespaces,
\normalsize{\texttt{\arxivref{1610.09378}{arXiv:1610.09378}}}\ignorespaces
\bibitem{Cuomo:2017wme}
G.~F. Cuomo, D.~Karateev \& P.~Kravchuk,
\textit{``{General Bootstrap Equations in 4D CFTs}''},
\doiref{10.1007/JHEP01(2018)130}{JHEP \textbf{1801}, 130
  (2018)\ignorespaces}\ignorespaces,
\normalsize{\texttt{\arxivref{1705.05401}{arXiv:1705.05401}}}\ignorespaces,
[,57(2017)]\ignorespaces
\bibitem{Cvitanovic:2008zz}
P.~Cvitanovic,
\textit{``{Group theory: Birdtracks, Lie's and exceptional groups}''},
Princeton,~USA:~Univ.~Pr.~(2008)~273~p \textbf{1801}, P.~Cvitanovic
  (2008)\ignorespaces\ignorespaces,
\href{http://press.princeton.edu/titles/8839.html}{\texttt{http://press.princeton.edu/titles/8839.html}}
\bibitem{Costa:2014rya}
M.~S. Costa \& T.~Hansen,
\textit{``{Conformal correlators of mixed-symmetry tensors}''},
\doiref{10.1007/JHEP02(2015)151}{JHEP \textbf{1502}, 151
  (2015)\ignorespaces}\ignorespaces,
\normalsize{\texttt{\arxivref{1411.7351}{arXiv:1411.7351}}}\ignorespaces
\bibitem{Costa:2016hju}
M.~S. Costa, T.~Hansen, J.~Penedones \& E.~Trevisani,
\textit{``{Projectors and seed conformal blocks for traceless mixed-symmetry
  tensors}''},
\doiref{10.1007/JHEP07(2016)018}{JHEP \textbf{1607}, 018
  (2016)\ignorespaces}\ignorespaces,
\normalsize{\texttt{\arxivref{1603.05551}{arXiv:1603.05551}}}\ignorespaces
\bibitem{Karateev:2017jgd}
D.~Karateev, P.~Kravchuk \& D.~Simmons-Duffin,
\textit{``{Weight Shifting Operators and Conformal Blocks}''},
\doiref{10.1007/JHEP02(2018)081}{JHEP \textbf{1802}, 081
  (2018)\ignorespaces}\ignorespaces,
\normalsize{\texttt{\arxivref{1706.07813}{arXiv:1706.07813}}}\ignorespaces,
[,91(2017)]\ignorespaces
\bibitem{Costa:2018mcg}
M.~S. Costa \& T.~Hansen,
\textit{``{AdS Weight Shifting Operators}''},
\doiref{10.1007/JHEP09(2018)040}{JHEP \textbf{1809}, 040
  (2018)\ignorespaces}\ignorespaces,
\normalsize{\texttt{\arxivref{1805.01492}{arXiv:1805.01492}}}\ignorespaces
\bibitem{Rejon-Barrera:2015bpa}
F.~Rejon-Barrera \& D.~Robbins,
\textit{``{Scalar-Vector Bootstrap}''},
\doiref{10.1007/JHEP01(2016)139}{JHEP \textbf{1601}, 139
  (2016)\ignorespaces}\ignorespaces,
\normalsize{\texttt{\arxivref{1508.02676}{arXiv:1508.02676}}}\ignorespaces
\bibitem{Streater:1989vi}
R.~F. Streater \& A.~S. Wightman,
\textit{``{PCT, spin and statistics, and all that}''},
Redwood~City,~USA:~Addison-Wesley~(1989)~207~p.~(Advanced~book~classics)
  \textbf{1601}, A.~S. Wightman (1989)\ignorespaces\ignorespaces
\bibitem{Fradkin:1996is}
E.~S. Fradkin \& M.~{\relax Ya}. Palchik,
\textit{``{Conformal quantum field theory in D-dimensions}''},
Dordrecht,~Netherlands:~Kluwer~(1996)~461~p.~(Mathematics~and~its~applications.~376)
  \textbf{1601}, M.~{\relax Ya}. Palchik (1996)\ignorespaces\ignorespaces
\bibitem{Rychkov:2016iqz}
S.~Rychkov,
\textit{``{EPFL Lectures on Conformal Field Theory in D>= 3 Dimensions}''},
\normalsize{\texttt{\arxivref{1601.05000}{arXiv:1601.05000}}}\ignorespaces
\end{thebibliography}

\end{document}